\documentclass[%
 reprint,
superscriptaddress,
nofootinbib,
amsmath,amssymb,
 aps,
prl,
]{revtex4-2}
\usepackage[T1]{fontenc}

\usepackage[colorlinks]{hyperref}
\hypersetup{colorlinks=true,
	linkcolor=red,
	anchorcolor=black,
	citecolor=blue,
	urlcolor=black
}
\usepackage{xr}

\usepackage[utf8]{inputenc}

\usepackage{amsfonts}
\usepackage{amssymb}
\usepackage{mathrsfs}
\usepackage{amsthm}
\usepackage{amssymb}
\usepackage{float}
\usepackage{graphicx}
\usepackage{dcolumn}
\usepackage{bm}
\usepackage{xcolor}
\usepackage{natbib}
\usepackage[capitalise]{cleveref}

\providecommand{\be}{\begin{equation}}
\providecommand{\ee}{\end{equation}}

\newcommand{\bsi}{\boldsymbol{s}}%
\newcommand{\bsG}{\boldsymbol{G}}%
\newcommand{\bstheta}{\boldsymbol{\theta}}%
\newcommand{\bM}{\boldsymbol{M}}%
\newcommand{\bsx}{\boldsymbol{x}}%
\newcommand{\bsv}{\boldsymbol{v}}%
\newcommand{\bsm}{\boldsymbol{m}}%
\newcommand{\bsM}{\boldsymbol{M}}%
\newcommand{\bscalM}{\boldsymbol{\mathcal{M}}}%
\newcommand{\bsJ}{\boldsymbol{J}}%
\newcommand{\bsh}{\boldsymbol{H}}%
\newcommand{\nn}{\nonumber}%
\newcommand{\delays}{\tau_{\rm M}}%
\newcommand{\maxtimes}{N_{\rm T}}%
\newcommand{\avtimes}{\maxtimes-\delays}%
\newcommand{\anorm}{\frac{1}{\avtimes}}%
\newcommand{\dims}{D}%
\newcommand{\dt}{\delta t}%
\newcommand{\autocorr}{R}%
\newcommand{\akaike}{\textrm{AIC}}%

\newcommand{\tsumnorm}[1]{\anorm\sum_{#1=\delays+1}^{\maxtimes}}%
\newcommand{\tsum}[1]{\sum_{#1=\delays+1}^{\maxtimes}}%
\newcommand{\tprod}[1]{\prod_{#1=\delays+1}^{\maxtimes}}%
\newcommand{\dsum}[1]{\sum_{#1=1}^{\delays}}%
\newcommand{\tavg}[1]{\left\langle #1 \right\rangle _{t}}%

\begin{document}

\setlength{\parskip}{5pt plus 0pt minus 0pt}

\preprint{APS/123-QED}

\title{Inverse modeling of time-delayed interactions via the dynamic-entropy
formalism}

\author{Elena Agliari}
\affiliation{Dipartimento di Matematica, Sapienza Universit\`a di Roma, Roma, Italy}

\author{Francesco Alemanno}%
\affiliation{Dipartimento di Matematica e Fisica, Universit\`a del Salento, Lecce,  Italy}%

\author{Adriano Barra}
\email{adriano.barra@uniroma1.it}
\affiliation{Dipartimento di Scienze di Base e Applicate per l'Ingegneria , Sapienza Universit\`a di Roma, Roma, Italy}%
\affiliation{Istituto Nazionale di Fisica Nucleare, Sezione di Lecce, Italy}

\author{Michele Castellana}
\affiliation{Laboratoire PhysicoChimie, Institut Curie, CNRS UMR168, Paris, France}%

%
\author{Daniele Lotito}
\affiliation{Dipartimento di Informatica, Università di Pisa, Pisa, Italy} 
\author{Matthieu Piel}
\affiliation{Laboratoire de Biologie Cellulaire et Cancer, Institut Curie, CNRS UMR168, Paris, France} 
%
%
%


\date{\today}

\begin{abstract}
Although instantaneous interactions are unphysical,  a large variety of maximum entropy statistical inference methods  match  the model-inferred and the empirically-measured equal-time correlation functions. Focusing on collective motion of active units, this constraint  is reasonable when the interaction timescale  is much faster than that of the interacting units,  as in starling flocks, yet it fails in a number of counter examples, as in leukocyte coordination (where  signalling proteins diffuse among  two cells). Here, we relax this assumption and develop a path integral approach to maximum-entropy framework, which includes delay in signalling. Our method  is able to infer the strength of couplings and fields, but also the time required by the couplings to completely transfer information among the  units. We demonstrate the validity of our approach providing excellent results on synthetic datasets of non-Markovian trajectories generated by the Heisenberg-Kuramoto and Vicsek models equipped with delayed interactions. As a proof of concept, we also apply the method to  experiments on dendritic migration, where matching equal-time correlations results in a significant information loss.  
\end{abstract}

\pacs{Valid PACS appear here}
\maketitle

The ability to move is a fundamental property of many living systems, ranging on different scales, from animals to cells \cite{Locomotion2018}. As for the latter, advances in live imaging allowed scientists to collect big data for which statistical analysis has now become a robust and reliable tool \cite{Meijering-2012}. These investigations suggest that cells often migrate in groups and communicate as they move.  Establishing whether such interactions are present is, in many cases, of utmost importance to possibly control and anticipate the evolution of a system.
However, answering this question is notoriously difficult, especially when unit-to-unit communication is not supported by the existence of physical bonds, as in 
juxtacrine interactions, but  it is rather induced by some signaling pathways, 
as in paracrine interactions.
To inspect these possible interactions, several methods have been introduced in the past few decades. Among these, inferential techniques have been designed to take as input some  tracks of  moving units, such as birds in flocks or migrating cells, and output the parameters which provide information about the motion and the  existence of interactions among the units, and/or between the units and an external source, such a predator for a bird flock \cite{Procaccini2011}, or a cancer cell producing signalling proteins for leukocytes \cite{Alemanno2023}.

In particular, approaches based on the maximum entropy (ME) principle  \cite{jaynes1957information} have been proven successful in a broad variety of contexts, revealing interaction patterns resulting from amino-acid sequences in protein families \cite{seno2008maximum}, interaction structures of genetic networks \cite{lezon2006using}, effective interactions in networks of neurons \cite{bialek2012statistical}, etc. The state of the art of this inverse modeling still exhibits some limitations which  make it unsatisfactory  at work on units coordination for collective motion.
Specifically, this approach usually forces the theoretical and experimental equal-time correlation functions to match, implicitly assuming instantaneous interactions among units, however, this  assumption may not hold when the interaction is mediated by some chemicals which  take some time to diffuse from the emitter to the receiver. For example,  in cell-migration experiments, cell $i$ may locally release a chemical compound along its migratory path, and cell $j$ may cross the former path of $i$ at a later time and thus feel a delayed interaction with $i$ mediated by this compound. In this work, we introduce, develop and test a ME framework that allows unveiling not only time-lagged effective interactions but also, and more generally, interactions that arise indirectly as cell $i$ alters the environment of cell $j$.

Let us consider the motion of $N$ particles (i.e. units) in a $\dims$-dimensional space, and denote their trajectories by 
\begin{equation} 
\left\{ \bsx_i^t\right\}, \, ~~ 1 \leq  i \leq N, ~~~1 \leq t \leq  \maxtimes+1,
\end{equation}
where $\bsx_i^t$ is the $D$-dimensional position vector of particle $i$ at time $t$, and the superscript index indicates instants of time separated by $\dt$, for a total of $\maxtimes+1$ temporal samples.
The velocity $\bsv_{i}^{t}$ and direction of motion $\bsi_{i}^{t}$ of each particle $i$ at time $t$ then read, respectively, 
\begin{equation}\label{velocity}
\bsv_{i}^{t}\equiv\frac{\bsx_{i}^{t+1}-\bsx_{i}^{t}}{\dt},\quad\bsi_{i}^{t}\equiv\frac{\bsv_{i}^{t}}{\left| \bsv_{i}^{t}\right| } \ \text{and} \   \bsm_t \equiv \frac{1}{N}\sum_{i=1}^{N}\bsi_{i}^{t}
\end{equation}
is the instantaneous mean alignment of the moving units at time $t$ 
which shall also be referred to as the `magnetization', while  $\boldsymbol s^t \equiv \{\boldsymbol s_1^t, ..., \boldsymbol s_N^t \}$ is interpreted as a configuration (at time $t$) of $N$ Heisenberg spins.
For practical purposes,  we assume that the motion of these units may be temporally correlated up to a maximum temporal window of length $\delays$, with $0\le\delays<\maxtimes$. Hereafter we assume that the correct value for $\delays$ is known a priori; later, once the necessary tools are developed, we will provide a means for determining $\delays$ from the data in a self-consistent manner via the Akaike information criterion (AIC) \cite{aic_banks2017}.  

The two natural observables we use to quantify this temporal correlation are the temporally averaged magnetization $\bM$
\begin{equation}
\bM \equiv\tsumnorm t\bsm_t, \label{eq:tre}
\end{equation}
and the two-point correlation function $R$ with delay $\tau$
\begin{equation}\label{eq:quattro}
\autocorr\left(\tau \right) \equiv \tsumnorm t\,\bsm_{t} \cdot \bsm_{t-\tau}, ~1 \leq  \tau  \leq \delays.
\end{equation}
We seek $P(\bsi)$ as the minimal probability measure such that the theoretical averages for the magnetization \eqref{eq:tre} and correlation function \eqref{eq:quattro} match their empirical values and, accordingly, we obtain it following the ME principle \cite{jaynes1957information,agliari2020a,cavagna2014dynamical}: we define the path entropy $S$ as
\begin{equation}\label{ShannonS}
S\left[P\right]=-\int\mathcal{D}\bsi P(\bsi)\log P(\bsi)
\end{equation}
where 
$\mathcal{D}\bsi\equiv\prod_{t=1}^{\maxtimes}\prod_{i=1}^{N}d\bsi_{i}^{t}$ and $d\bsi_{i}^{t}$ is the surface
element of the $(\dims-1)$-dimensional sphere. 
%
\newline
To solve this optimization task, we introduce the Lagrangian multipliers $\boldsymbol{J} \equiv \{J_{\tau} \}_{\tau=1}^{\delays}$ and $\bsh \equiv \{H_{\ell}\}_{\ell=1}^D$, and obtain
\begin{small}
\begin{equation} \label{eq:sei}
P(\bsi)= 
\frac{1}{Z}\exp\left[\tsum t\sum_{i=1}^{N}\bsi_{i}^{t}\cdot\left(\frac{1}{N}\dsum{\tau}J_{\tau}\sum_{j=1}^{N}\bsi_{j}^{t-\tau}+\bsh\right)\right],
\end{equation}
\end{small}
\noindent
where $Z\equiv\int\mathcal{D}\bsi P[\bsi]$ is a normalization constant, see  the Appendix for details.  

Remarkably, $P(\bsi)$ can be looked at as a Boltzmann-Gibbs distribution and, consistently, the argument of its exponential  can be interpreted as a Hamiltonian. Thus, the Lagrange multipliers $\bsh$ and $J_{\tau}$ play, respectively, as an effective external field and as an effective interaction-strength, such that the direction of the $i$th unit at time $t$ tends to align with the direction of $\bsh$ and to align (misalign) with the average unit's direction at time $t - \tau$ if $J_{\tau}>0$ ($J_{\tau}<0$).
Having obtained an analytical expression for $P(\bsi)$, we  infer the parameters $\boldsymbol{J}$ and $\boldsymbol{H}$ from datasets.


Before proceeding we note that, in the following, the theoretical average over $P(\bsi)$ shall be denoted by brackets, while the experimental average shall be denoted by the superscript {\em E} (e.g. $\langle  \bsm_t \rangle$  and $\bsm^{\rm E}_t$).
\newline
As averaging over the measure $P(\bsi)$ is computationally cumbersome (as we have to integrate over $N_T \times N$ coupled variables at once),   {\em en route} toward an approach that can be routinely applied in a lab, hereafter we explore an approximation, whose consistency can be checked a posteriori. First, in \cref{eq:sei} we replace the internal field with its empirical counterpart
\be\label{eq:sette}
\frac{1}{N}\dsum{\tau}J_{\tau}\sum_{j=1}^{N}\bsi_{j}^{t-\tau} \rightarrow \dsum{\tau}J_{\tau}\bsm_{t-\tau}^{\rm E}.
\ee 
The resulting distribution is referred to as $P_{\rm A}$ to highlight that it is an approximation of $P$.
Denoting by $\langle\rangle_{\rm A}$ the related average, we get
%
%
\begin{align}
\label{eq:otto}
\langle\bsm_{t}\rangle_{\rm A}&=\int \left( \prod_{t=\delays+1}^{\maxtimes}\prod_{i=1}^{N}d\bsi_{i}^{t} \right) P_{\rm A}[\bsi] \frac{1}{N}\sum_{i=1}^{N}\bsi_{i}^{t} =\nn\\
&=\begin{cases}
\bscalM\left(\dsum{\tau}J_{\tau}\bsm_{t-\tau}^{\rm E}+\bsh\right) & \textrm{ if } \delays\le t\le\maxtimes,\\
\bsm_{t}^{\rm E} & \textrm{ if } 1\le t\le\delays,
\end{cases}
\end{align}
where 
\begin{equation}
\bscalM(\bsx)\equiv\frac{\bsx}{\left| \bsx\right| }\frac{\mathcal{I}_{1}\left(\left| \bsx\right| \right)}{\mathcal{I}_{0}\left(\left| \bsx\right| \right)}, 
\end{equation}
and $\mathcal{I}_{0},\mathcal{I}_{1}$ are hyperbolic Bessel functions of order $0$ and $1$, respectively \cite{abramowitz1965handbook}. 
    
Next, we notice that $P_{\rm A}$ is factorized with respect to every variable $\bsi_{i}^{t}$, hence $\langle\bsm_{t}\bsm_{t^{\prime}}\rangle_{\rm A}=\langle\bsm_{t}\rangle_{\rm A}\langle\bsm_{t^{\prime}}\rangle_{\rm A}$.
This allows us to recast the constraints in the ME extremization as
\begin{footnotesize}
\begin{eqnarray}
\label{eq:dieci}
 \bM^{\rm E} &=& \tsumnorm t\langle\bsm_{t}\rangle_{\rm A},\\
\label{eq:undici}
\autocorr^{\rm E}(\tau) &=& \tsumnorm t\langle\bsm_{t}\rangle_{\rm A}\cdot\langle\bsm_{t-\tau}\rangle_{\rm A},~~  1 \le \tau \le \delays.
\end{eqnarray}
\end{footnotesize}
The non-linearities in $\boldsymbol J$ and $\boldsymbol H$ introduced by the operator $\bscalM$ make the direct evaluation of these parameters from Eqs.~\eqref{eq:dieci}-\eqref{eq:undici} non-feasible.  To circumvent this difficulty, we can note that  these equations  are simultaneously fulfilled by requiring that
\begin{equation}
\label{eq:dodici}
\langle\bsm_{t}\rangle_{\rm A}=\bsm_{t}^{\rm E},\qquad  \delays + 1 \leq t \leq \maxtimes,
\end{equation}
which, exploiting \cref{eq:otto}, becomes
\begin{equation}\label{eq_2}
\bscalM\left(\dsum{\tau}J_{\tau}\bsm_{t-\tau}^{\rm E}+\bsh\right)=\bsm_{t}^{\rm E},\qquad \delays +1 \leq t \leq \maxtimes,
\end{equation}
and, finally, by introducing the inverse function $\bsG_{t}^{\rm E}  \equiv  \bscalM^{-1}\left(\bsm_{t}^{\rm E}\right)$ we can shift the non-linearity on the experimentally-determined term and get
\begin{equation}
\dsum{\tau}J_{\tau}\bsm_{t-\tau}^{\rm E}+\bsh =  \bsG_{t}^{\rm E},\qquad \delays +1 \leq t \leq \maxtimes. \label{eq:quattordici}
\end{equation}
Hence, the  solution of the original integral over the past history of all the particles \eqref{ShannonS} is effectively approximated in terms of the solution of the linear system \eqref{eq:quattordici}.
\newline
We stress that \cref{eq:dodici} constitutes a stricter condition than Eqs. \eqref{eq:dieci} and \eqref{eq:undici}, because it stems from approximation \eqref{eq:sette} that removed degrees of freedom.  To avoid  that \eqref{eq:quattordici} is overconstrained, we include an additional and controllable term in it, i.e., 
$\dsum{\tau}J_{\tau}\bsm_{t-\tau}^{\rm E}+\bsh+\sigma\boldsymbol{\epsilon}_{t} =  \bsG_{t}^{\rm E}$, where $\boldsymbol{\epsilon}_{t} \sim  \mathcal{N}\left(\boldsymbol{0}_{\dims},\boldsymbol{1}_{\dims}\right)$ and $\sigma \ge 0$ is a further parameter to be determined: we have modified  the constraint in such a way that the relation \eqref{eq:quattordici} does not hold deterministically. Thus, 
the related log-likelihood $l_{\rm A}$ reads 
\begin{eqnarray}
\nonumber
l_{\rm A}(\bsJ,\bsh,\sigma |\bsi^{\rm E})&\equiv&-\frac{1}{2\sigma^{2}}\tsum t\left| \dsum{\tau}J_{\tau}\bsm_{t-\tau}^{\rm E}+ \bsh-\bsG_{t}^{\rm E}\right| ^{2}\\
&-&\left(\avtimes\right)\frac{\dims}{2}\log\left(2\pi\sigma^{2}\right).\label{eq:gaussian_loglikelihood}
\end{eqnarray}
We now  obtain the optimal estimates for $\bsJ, \bsh$ and $\sigma$. Setting $\nabla_{\bsh}l_{\rm A}=0$, we obtain
\begin{equation} \label{eq:estH}
\bsh=\langle \bsG_{t}^{\rm E}\rangle _{t}-\dsum{\tau}J_{\tau}\left\langle \bsm_{t-\tau}^{\rm E}\right\rangle _{t},
\end{equation}
where $\tavg{ \cdot }\equiv\tsumnorm t (\cdot)$ 
denotes the time average, and, proceeding analogously for  $\nabla_{J_{\tau}}l_{\rm A}=0$,  we get
\begin{equation} \label{eq:estJ}
\bsJ  =\boldsymbol{A}^{-1}\cdot\boldsymbol{B},
\end{equation}
with
\begin{eqnarray}
A_{\lambda\tau} & \equiv&  \left\langle \bsm_{t-\tau}^{\rm E}\cdot\bsm_{t-\lambda}^{\rm E}\right\rangle _{t}-\left\langle \bsm_{t-\tau}^{\rm E}\right\rangle _{t}\cdot\left\langle \bsm_{t-\lambda}^{\rm E}\right\rangle _{t},\\
B_{\lambda} & \equiv & \langle \bsG_{t}^{\rm E}\cdot\bsm_{t-\lambda}^{\rm E} \rangle _{t}-\langle \bsG_{t}^{\rm E}\rangle _{t}\cdot \langle \bsm_{t-\lambda}^{\rm E} \rangle _{t}.
\end{eqnarray}
Finally, setting $\nabla_{\sigma}l_{\rm A}=0$, we obtain
\begin{equation}
\sigma^{2}=\frac{1}{\dims}\left\langle \left| \bsh+\dsum{\tau}J_{\tau}\bsm_{t-\tau}^{\rm E}-\bsG_{t}^{\rm E}\right| ^{2}\right\rangle _{t}.\label{eq:venti}
\end{equation}

As detailed in the Appendix, 
an estimate of the errors for the inferred parameters can be obtained by using the Fisher information for the log-likelihood  \cite{Frieden-2004}. In fact, 
\begin{equation}
\textrm{Var}\left(\bsJ,\bsh,\sigma\right) \approx  \textrm{diag}\left(\boldsymbol{I}^{-1}\right),
\end{equation}
where the approximation holds as long as the sample size is large enough and $\boldsymbol{I}$  is the Fisher information matrix, which in the present setting reads
\begin{equation}
\nonumber
\boldsymbol{I} = \frac{\avtimes}{\sigma^{2}}\left(\begin{array}{ccc}
\left\langle \bsm_{t-\tau}^{\rm E}\cdot\bsm_{t-\lambda}^{E}\right\rangle _{t} & \left\langle \bsm_{t-\tau}^{\rm E}\right\rangle _{t} & 0\\
\left\langle \bsm_{t-\lambda}^{E}\right\rangle _{t} & {\boldsymbol \delta} & {\boldsymbol 0}\\
0 & {\boldsymbol 0} & 2\dims
\end{array}\right).
\end{equation}

So far in our analysis, we assumed that $\delays$ is known, yet in most practical cases it is not. To estimate $\delays$, we rely on the AIC \cite{aic_banks2017}, which provides a tool for model selection, each model being characterised by a different value of $\delays$.  In general, given a dataset and a model 
with a mean squared error $\epsilon^{2}$, $n_{\rm p}$ parameters and $n_{\rm o}$ observations, the quantity
\begin{equation}
\akaike\equiv\frac{2n_{\rm p}}{n_{\rm o}}+\log\left(\epsilon^{2}\right)\label{eq:ventidue}
\end{equation}
assesses the quality of such a model for the data available. In fact, the AIC favors models with a small error \emph{and} a small number of parameters.
In the analysis above, the parameters in \cref{eq:ventidue} are
\begin{equation}
\nonumber
n_{\rm o}  =  \dims\left(\maxtimes-\delays\right),~~~~ n_{\rm p}  =  \delays+\dims+1,~~~~ \epsilon^{2}  =  \frac{\maxtimes}{\maxtimes-\delays}\sigma^{2},
\end{equation}
thus
\begin{equation}
\akaike=2\frac{\delays+\dims+1}{D\left(\avtimes\right)}+\log\frac{\maxtimes ~ \sigma^{2}}{\maxtimes-\delays}.\label{eq:ventitre}
\end{equation}
In the following, after performing inference for multiple values of $\delays$
and determining $\sigma^2$ from \cref{eq:venti}, we evaluate \cref{eq:ventitre} for each inferred
model (i.e. by varying $\tau_M$) and we choose the effective $\tau_M$ as that with the smallest value of $\akaike$: the maximal (in modulus) $\boldsymbol J$ within the time window $0\leq t \leq \tau_M$ is then interpreted as the best inferred coupling.
\begin{figure}
\begin{centering}
\includegraphics[width=0.5\textwidth]{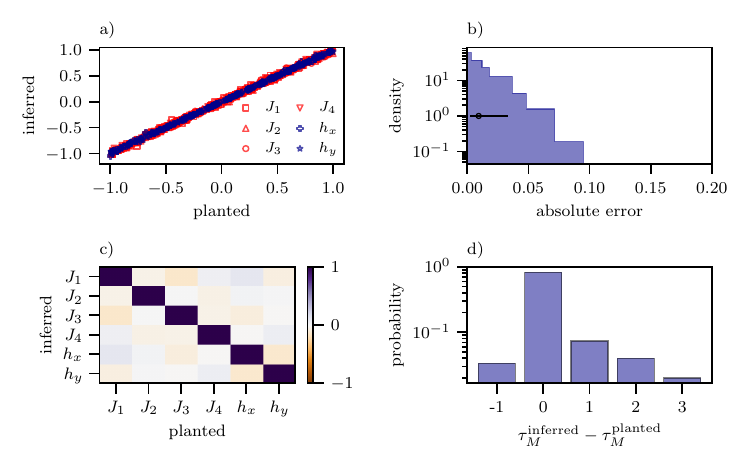}
\par\end{centering}
\caption{\label{fig:me_vs_itself}  Validation of our inference method with the HKM. Here $D=2$, $\delays = 4$, $\sigma=0.1$, and we drew $150$ planted parameters $J_{1},\cdots,J_{\delays},H_{x},H_{y}$ independently and identically from a uniform distribution over the range $(-1,1)$. 
~a) Scatter plot of planted vs inferred parameters: they match along the diagonal as they should.
~b) Histogram of the absolute error between $\protect\bsJ,\protect\bsh$
planted and $\protect\bsJ,\protect\bsh$ inferred: for each realisation of the process we calculate $|\bsJ^{\rm inferred}-\bsJ^{\rm planted}|$ and $|\bsh^{\rm inferred}-\bsh^{\rm planted}|$ and the related $150 \times (D + \delays)$ components make up the sample; the black dot represents the average and the black horizontal line represents a confidence interval of $68 \%$; the average error is  $\sim 10^{-2}$, which is the expected amount of error given the size of the planted dataset. 
~c) Pearson correlation matrix between planted and inferred parameters: as expected distinct types of parameters, e.g.,  $\protect\bsh$ and $\protect\bsJ$, are uncorrelated. 
~d) Histogram of $\protect\delays^{\rm inferred}-\protect\delays^{\rm planted}$: the peak of the distribution correctly returns the inferred delay as the planted one. }
\end{figure}

We test our framework on both synthetic and real datasets. The models used to generate the synthetic datasets are the Heisenberg-Kuramoto model (HKM) \cite{Alemanno2023}, whose Boltzmann-Gibbs distribution is precisely \eqref{eq:sei}, and the (topological or metric \cite{VicsekDual}) Vicsek model (VM) \cite{vicsek1995novel}, both equipped with a (tunable) delay in the interactions \cite{DelayedVicsek}. Each model has its own set of parameters, which we denote as \textit{planted} parameters, while those returned by the ME extremization are labelled as {\em inferred}.
\newline
As a conditio sine qua non for further analyses, the validation of our ME method on the HKM is shown in Fig.~\ref{fig:me_vs_itself} providing excellent results, even away from the Markovian limit $\tau_M=1$.

\begin{figure}
\begin{centering}
\includegraphics[width=0.48\textwidth]{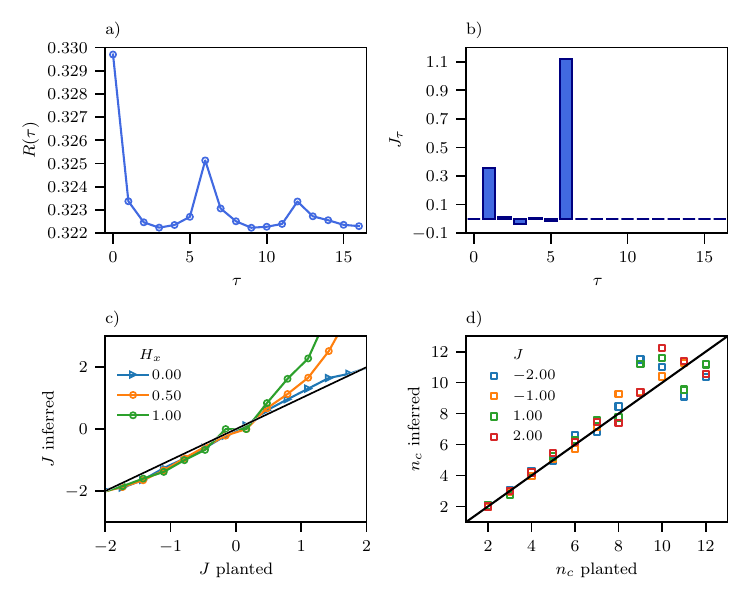} 
\par\end{centering}
\caption{\label{fig:Testing-VicsecTopA} Validation of our inference method with the topological VM, simulated in a periodic, two-dimensional square lattice with size $L=150$. $N = 100,~\Delta = 6,~\delta t=1,~v_0 = 1,~n_{\rm c}=4,~\bsh=(H_x, 0)$. $H_x$ and $J$ are i.i.d. over the set $[0,2]\times [0, 2]$. The analogous picture for the metric case is reported in the Appendix. Panels: a) example of the empirical correlation $R(\tau)$ vs $\tau$. b) Example of the (related) inferred coupling: note the presence of two peaks, the former (at $\tau=1$) is the Markovian self-interaction, the latter (at $\tau=\Delta=6$) is the non-Markovian contribution by nearest neighbors, in accordance with \eqref{eq:ventiquattro}, \eqref{eq:venticinque}. c) Scatter plot of the inferred vs the planted values of the coupling: note that the larger the (external) field $H_x$, the greater the overestimation of the (inferred) coupling. d) Scatter plot of the number of nearest neighbors $n_c$ estimated as the ratio between the intensities of the coupling at $\tau=1$ (the Markovian self-interaction accounting for inertia) and at $\tau=6$ (the non-Markovian interaction with other units), i.e., $n_c = 1 + J_{\Delta}/J_{1}$.}
\end{figure}

\begin{figure}
\begin{centering}
\includegraphics[width=0.48\textwidth]{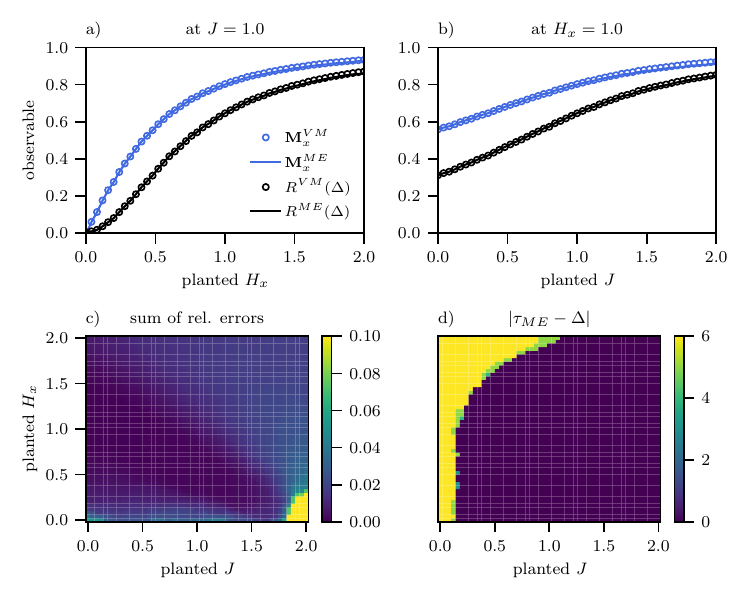} 
\par\end{centering}
\caption{\label{fig:Testing-VicsecTopB} Validation of our inference method with the topological VM whose setting is the same as in Fig.~\ref{fig:Testing-VicsecTopA}. The superscripts $V$ and $ME$ pertain to quantities calculated for the VM or inferred via the ME method, respectively.  Panels: a) Observables $\bsM$ and $R(\Delta)$ as functions of $H_x$, with $J = 1.0$. b) Observables $\bsM$ and $R(\Delta)$ as functions of $J$, with $H_x = 1.0$.
c) Sum of the relative errors $\frac{|M^{ME}-M^{VM}|}{|M^{VM}|}+\frac{|R^{VM}(\Delta)-R^{ME}(\Delta)|}{R^{VM}(\Delta)}$ as functions of $H_x$ and $J$, 
d) Absolute deviation between the inferred $\tau_{ME} := \text{argmax}_{\tau}|J_{ME}(\tau)|$ and the planted parameter $\Delta$ as a function of $J$ and $H_x$.
The analogous figure for the metric case is reported in the Appendix. }
\end{figure}

Next, introducing the i.i.d. noise $\boldsymbol{\eta}_{k}^{t}\sim\mathcal{N}(\boldsymbol{0},\boldsymbol{1})$ and using $\Delta$ to account for the planted delay in signalling,  we consider the  VM in a $2D$ square lattice of size $L$ with periodic boundary conditions, defined by the following dynamical equations \cite{vicsek1995novel,ginelli2010relevance}
\begin{eqnarray}\label{eq:ventiquattro}   
\bsv_{k}^{t+1}&=&v_{0}\boldsymbol{\Gamma}\Bigg(\!\frac{J}{\left|n_{k}^{t}\right|}\sum_{j\in n_{k}^{t}}\!\!\frac{\bsv_{j}^{t-D_{ik}}}{v_{0}}+\!\bsh\!+\!\sqrt{\delta t}\,\boldsymbol{\eta}_{k}^{t}\!\Bigg),\\  
\label{eq:venticinque}
\bsx_{k}^{t+1}&=&\bsx_{k}^{t}+\delta{t}\,\bsv_{k}^{t+1} ~~~k=1,\cdots,N.
\end{eqnarray}
where $D_{ij}=(\Delta-1) (1-\delta_{i}^{j})$ while  $\bsv_{i}^{t}$, $\bsx_{i}^{t}$  are the velocity and position of the $i${th} particle at time $t$, respectively, and two temporally consecutive frames are distant $\delta t$. 
In the topological implementation of the VM, $n_k^{t}$ is the set of topological neighbors of the $k$th particle (including the $k$th particle itself) and $|n_{k}^{t}| := n_{c}$ is the number of neighbors of the $k$th particle at time $t$. In the metric counterpart,  deepened in the Appendix, $n_{k}^{t}$ is the set of the closer units within an interaction range.
As standard in the VM, $\boldsymbol{\Gamma}(\bsv) \equiv \bsv/\left| \bsv\right| $, thus the velocity of each particle has constant magnitude $v_{0}$.
Finally, the magnetization of the VM is given by  
\begin{equation}
\bsm_t=\frac{1}{N}\sum_{k=1}^{N}\boldsymbol{\Gamma}(\bsv_{k}^{t}),
\end{equation}
which is the adaptation of the last expression in \cref{velocity} to the case. 
In the validation of the ME inference for the VM we  set $H_y = 0$: the parameters to estimate are thus $J, H_x$ and the delay (whose inferred value we call $\tau_{ME}$ as opposite to $\Delta$ that is the planted value) as reported in Fig.~\ref{fig:Testing-VicsecTopA}. 
Further, to deepen the statistical properties of the inferred model, we also compare  the lowest-order  moments of some observables for trajectories generated by  the inferred ME model and  by the planted VM: in particular, we measure the magnetization $\bsM$ and the correlation function $R(\tau=\Delta)$ to check that their values are statistically compatible: Fig.~\ref{fig:Testing-VicsecTopB} provides a 
positive test of the inference method for the topological VM (and the same holds also for the metric case, discussed in the Appendix). 

\begin{figure}[tb]
\centering{}\includegraphics[width=0.48\textwidth]{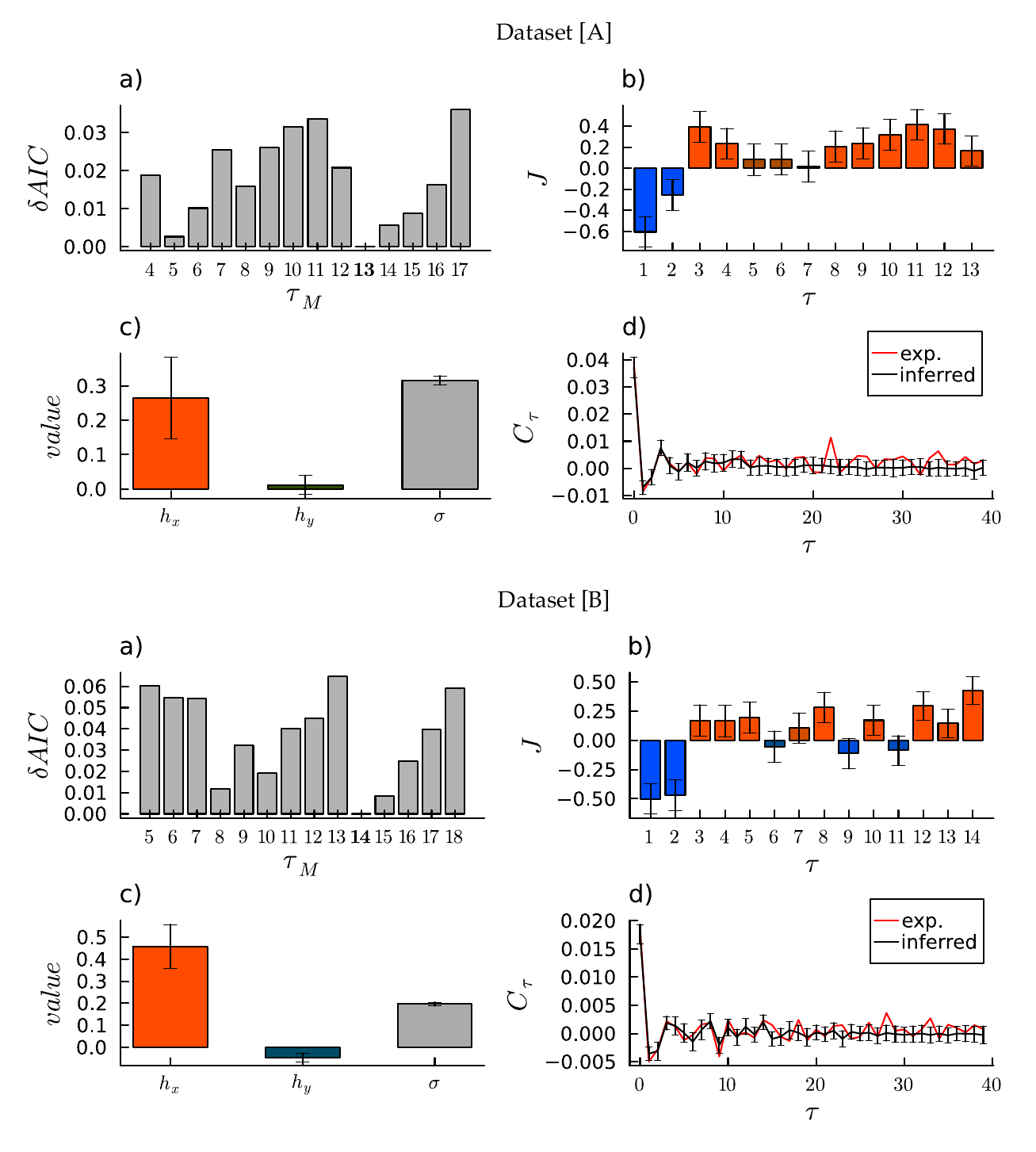}
\caption{\label{fig:inference} Inference for the chemokine-gradient dataset ($\delta t = 2$ min), for both regions A and B. a) Difference between the AIC and its minimal value,  as a function of the total number of interactions $\delays$
b) Inferred  values of the delayed interaction $J_{\tau}$ and  corresponding errors.
c) Inferred values of the two  components of the field, $H_x$ and $H_y$, 
and inferred value of $\sigma$. d) Correlation function $C(\tau)$, see \cref{eq:ventisette}. }
\end{figure}

Finally, to highlight the inconsistency  that would result by neglecting interaction delays in real experiments, we consider datasets on dendritic migration in a chemokine gradient \cite{agliari2020a}: all the aspects of biological relevance will be discussed elsewhere, here we just focus on the methodology. 
Positions of all cells were recorded at (ordered) multiple times and we studied two regions of the experiment, the former (region A) where the gradient is low and the latter (region B) where the gradient is high,  providing two datasets suitable to be addressed by our approach, see the Appendix for details. 
We estimate
$$
\delta\akaike(\delays) \equiv \akaike(\delays)-\min_{\delays'}\akaike(\delays'),
$$ 
i.e., the difference between  $\akaike$ for a given $\delays$ and  $\akaike$ for the optimal $\delays$,  the inferred parameters $J_{\tau}$, $\bm H$,  
$\sigma$ and  the connected correlation function 
\begin{equation}\label{eq:ventisette}
C(\tau) \equiv \tavg{\bsm(\bsi^{t})\cdot\bsm(\bsi^{t-\tau})} - \tavg{\bsm(\bsi^{t})}\cdot\tavg{\bsm(\bsi^{t-\tau})}
\end{equation}
for the optimal value of $\delays$.
These results are reported in \cref{fig:inference}  for both datasets A and B, highlighting marked  differences between the prediction by the standard ME and our technique. Indeed, as shown in \cref{fig:inference}a, the optimal $\tau$ is $13$ for region $\rm A$ and $14$ for region $\rm B$, and,  as shown in \cref{fig:inference}b,  the corresponding couplings $J_{\tau=13}$ and $J_{\tau=14}$ are both positive, while in the $\tau \to 0$ part of these panels the interactions are actually repulsive: accordingly the connected correlation functions provided in \cref{fig:inference}d are small but systematically different from zero for $\tau<\tau_M$, then they vanish. As expected the method correctly returns a higher field in region $B$, w.r.t. $A$, as shown in \cref{fig:inference}c. A further last control test on this inference can be achieved simply by reshuffling the frames and feeding the ME method with this random permutation: in this case the coupling should disappear yet the field should be preserved and, indeed, this is the case, as shown in the Appendix (see Fig. \ref{fig:NOinference}).
\newline
\newline
To summarize, we revised the standard ME method to encompass scenarios where communication among units is slow: the dependence over the whole history of any particle has to be considered and the emerging dynamics does not have to be any longer Markovian. While from a gnoseological perspective such general setting should always be preferable to versions where equal-time correlators are constrained, when communication among units is reasonably faster than the fastest timescale of their dynamics,   standard ME and our method return the same outcome. Whatever the route, in a nutshell, we use the knowledge stemming from empirical correlations to construct an effective Hamiltonian model that correctly reproduces the observed motion (e.g. its statistical characterstics, at some prescribed order, depending on how much empirical information we relied upon and thus how many Lagrange multipliers we used). The interactions that we infer via maximum entropy should however be seen as effective interactions as, in principle, we do not know the model that generated the dynamics. For instance, when analysing trajectories generated by the Vicsek model, the details of the generating model do not enter in our analysis, nevertheless the inferred  values of these interactions -despite being {\em effective}- carry valuable information about the motion (furthermore underlying model may not even exist, and in general it is now known). 
%
\newline
\newline
We acknowledge financial support from PNRR MUR (PE0000013-FAIR), MAECI (F85F21006230001), PRIN MUR (20229T9EAT) and Sapienza University of Rome (RM12117A8590B3FA). 
\newline
The Authors are grateful to Andrea Cavagna, Irene Giardina, Theresa Jakuszeit  and Raphael Voituriez  for enlightening discussions.

\section{Appendix}

In this Appendix, we present details of  the mathematical framework developed for inferring delayed interactions among dynamical units, such as a group of migrating cells, using only their trajectories. First,  we present our approach to generalize the maximum entropy (ME) method, standardly based on matching equal-time correlation, in order to cope with delay in signalling, then  we validate our method on synthetic and real datasets. As for the synthetic ones, we focus on trajectories generated by the Heisenberg-Kuramoto model (HKM)\cite{Alemanno2023} with delay  and on trajectories generated according to the Vicsek model (VM)\cite{viscek+al_95}, both in its topological as well as metric implementations; finally,  we consider a real test-case focusing on the trajectories of dendritic cells migrating toward a chemoattractive source, namely in the presence of a chemokine gradient \cite{agliari2020a}.

\section{Dynamic maximum-entropy formalism\label{dme}}

\subsection{Problem formulation}\label{sec:intro}
Let us consider the motion of $N$ particles in a $\dims$-dimensional space, and denote by 
\begin{equation} 
\left\{ \bsx_i^t\right\}, \, ~~ 1 \leq  i \leq N, ~~~1 \leq t \leq  \maxtimes+1
\end{equation}
the set of trajectories, where $\bsx_i^t$ is the position of the particle $i$ at time $t$ and the superscript index indicates instants of time separated by $\dt$, for a total of $\maxtimes+1$ temporal samples.
By using these collected positions $\bsx_{i}^{t}$, we can evaluate
the velocity $\bsv$ and direction of motion $\bsi$ of each particle $i$ at time $t$:
\begin{equation}
\bsv_{i}^{t}\equiv\frac{\bsx_{i}^{t+1}-\bsx_{i}^{t}}{\dt},\quad\bsi_{i}^{t}\equiv\frac{\bsv_{i}^{t}}{\left| \bsv_{i}^{t}\right| }, \quad\,1 \leq  i \leq N, ~~~1 \leq t \leq  \maxtimes. \label{eq:vel_dir}
\end{equation}
Focusing on the collection of variables $\bsi\equiv\{\bsi_{i}^{t}\}, \,1 \leq  i \leq N, \,1 \leq t \leq  \maxtimes$, a natural framework to determine a probability measure for these units is the path-integral formalism  \cite{agliari2020a,cavagna2014dynamical}: we
define the path entropy $S$ as
\begin{equation}
S\left[P\right]=-\int\mathcal{D}\bsi P(\bsi)\log P(\bsi)
\end{equation}
where 
$\mathcal{D}\bsi\equiv\prod_{t=1}^{\maxtimes}\prod_{i=1}^{N}d\bsi_{i}^{t}$
is the integration measure and each $d\bsi_{i}^{t}$ is a surface
element of the $(\dims-1)$-dimensional sphere.
%
The average alignment of the moving units at time $t$ is given by 
\begin{equation}
  \bsm(\bsi^{t}) \equiv \frac{1}{N}\sum_{i=1}^{N}\bsi_{i}^{t},\label{eq_m}
\end{equation}
where
\be
\bsi^{t}\equiv\{\bsi_{i}^{t}\}, ~ 1 \leq i \leq N. 
\ee
The latter can be interpreted as a configuration of soft spins that evolve in time, thus, we will refer to $\bsm(\bsi^{t})$ as the magnetization at time $t$. 
Furthermore, to quantify the direction along which the units move on average, we introduce the temporally averaged magnetization $\bM$:
\begin{equation}
\bM(\bsi)\equiv\tsumnorm t\bsm(\bsi^{t}).\label{eq:bM}
\end{equation}
In our analysis, we assume that the motion of the units above can be temporally correlated up to a maximum temporal window whose length is $\delays$ with $0\le\delays<\maxtimes$. A natural observable to quantify this persistence is the two-point correlation function $R$ with delay $\tau$
\begin{equation}\label{eq_c_tau}\small
\autocorr\left(\tau,\bsi\right) \equiv\tsumnorm t\,\bsm\left(\bsi^{t}\right)\cdot\bsm\left(\bsi^{t-\tau}\right), ~~1 \leq  \tau  \leq \delays.
\end{equation}
For the moment we suppose that the correct value for $\delays$ is known a priori; in the following sections, after having developed the necessary tools, we also provide a recipe for selecting $\delays$ in a self-consistent manner for any dataset.
Notice that the expression in \eqref{eq_c_tau} implements a ``mean-field'' correlation as the direction of each unit $i$ is related to the direction of any other unit up to $\delays$ time steps before, regardless of their spatial distance.

Following the ME principle \cite{jaynes1957information}, we seek $P$ as the minimal probability measure whose average magnetization \eqref{eq:bM} and two-point correlation function \eqref{eq_c_tau} match their respective empirical values:
\begin{align}
 & \max_{P}S[P]\label{entropy_problem1}\\
 & \textrm{subject to}  \nn\\
 &   \int\mathcal{D}\bsi P(\bsi)\bM(\bsi)=\bM^{\rm E},\label{entropy_problem2}\\
 &  \int\mathcal{D}\bsi P(\bsi)\autocorr\left(\tau,\bsi\right)=\autocorr^{\rm E}(\tau), ~  1 \leq  \tau  \leq \delays,\label{entropy_problem3}\\
  &  \int\mathcal{D}\bsi P(\bsi)=1,\label{entropy_problem4}
\end{align}
where $\bM^{\rm E}$ and $\autocorr^{\rm E}$ are the empirical values of
the average magnetization and two-point correlation, respectively.

To solve \cref{entropy_problem1}, we use the method of Lagrangian
multipliers. The Lagrangian $S^{\star}$ reads 
\begin{equation}
  \begin{split}S^{\star}\left[P,J,\bsh\right]= &
    -\frac{1}{N(\avtimes)}\int\mathcal{D}\bsi P[\bsi]\log P[\bsi]+\\
 & +\dsum{\tau}J_{\tau}\left(\int\mathcal{D}\bsi P[\bsi]\autocorr\left(\tau,\bsi\right)-\autocorr^{\rm E}(\tau)\right)+\\
 & +\bsh\cdot\left(\int\mathcal{D}\bsi
   P(\bsi)\bM(\bsi)-\bM^{\rm E}\right)+\\
 & +\frac{\zeta}{N(\avtimes)}\left(\int\mathcal{D}\bsi P[\bsi]-1\right).
\end{split}
\end{equation} 
where $\boldsymbol{J} \equiv \{J_{\tau} \}_{\tau=1}^{\delays},  \bsh \equiv \{H_{\ell}\}_{\ell=1}^D, \zeta$ play as Lagrangian multipliers and the normalization factor $\frac{1}{N(N_T - \tau_M)}$ ensures that $S^{\star}$ is an intensive quantity with respect to $N$ and $N_T$, and that the final expression for $P(\bsi)$ is well defined (vide infra).  
The Lagrangian can be easily extremized with respect to $P$:
\begin{eqnarray}\nonumber
\frac{\delta S^{\star}}{\delta P}&=&-\frac{1}{N(\avtimes)}(\log P(\bsi)+1-\zeta) \\
&+&\dsum{\tau}J_{\tau}\autocorr\left(\tau,\bsi\right)+\bsh\cdot\bM(\bsi)=0
\end{eqnarray}
yielding
\begin{equation}\small
P(\bsi)=\frac{1}{Z}\exp\left\{ N(\avtimes)\left[\dsum{\tau}J_{\tau}\autocorr\left(\tau,\bsi\right)+\bsh\cdot\bM(\bsi)\right]\right\}, \label{eq:Prob_raw}
\end{equation}
where the Lagrangian multipliers $\bsJ,\bsh$  are determined implicitly by \cref{entropy_problem2,entropy_problem3,entropy_problem4}, and $Z \equiv \exp\{(1-\zeta)/[N(\avtimes)]\}$. 
By substituting \cref{eq:bM,eq_c_tau} into \cref{eq:Prob_raw},
we get
\begin{equation}\label{eq1}\small
P(\bsi)=\frac{1}{Z}\exp\left[N\tsum t\bsm\left(\bsi^{t}\right)\cdot\left(\dsum{\tau}J_{\tau}\,\bsm\left(\bsi^{t-\tau}\right)+\bsh\right)\right].
\end{equation}
Finally, by plugging \cref{eq_m} into \cref{eq1}, we obtain
\begin{equation}\small
P(\bsi)=\frac{1}{Z}\exp\left[\tsum t\sum_{i=1}^{N}\bsi_{i}^{t}\cdot\left(\frac{1}{N}\dsum{\tau}J_{\tau}\sum_{j=1}^{N}\bsi_{j}^{t-\tau}+\bsh\right)\right],\label{eq_total_prob}
\end{equation}
where the normalization constant $Z$ reads 
\begin{equation}\label{NormaliZZa}\small
Z=\int\mathcal{D}\bsi\exp\left[\tsum t\sum_{i=1}^{N}\bsi_{i}^{t}\cdot\left(\frac{1}{N}\dsum{\tau}J_{\tau}\sum_{j=1}^{N}\bsi_{j}^{t-\tau}+\bsh\right)\right].
\end{equation}
We emphasize that the resulting $P(\bsi)$ can be interpreted as a Boltzmann-Gibbs distribution whose exponential  defines an effective energy function that, consistently with thermodynamic principles, scales linearly with the number of degrees of freedom $N$ and $N_T$. More generally, the structure of $P(\bsi)$ suggests that the direction of the $i$th particle at time $t$
tends to be aligned with the direction given by $\bsh$ -- which plays as an external field -- and that, if $J_{\tau}>0$ (resp. $J_{\tau}<0$), the direction of the $i$th particle at time $t$ tends to be aligned (resp. misaligned) with the average unit's direction at time $(t - \tau)$ -- which plays as an internal field.

\subsection{Solution method}\label{sec:LogL}

In this section we present our strategy to obtain an estimate for the parameters $\bsJ$ and $\bsh$, starting from a sample of experimental data. We first recast the problem into a maximum likelihood setting, which allows us to implement a simplification in the expression for $P(\bsi)$, yielding to an approximated expression denoted as $P_A(\bsi)$. This approximation is made necessary by the prohibitive difficulty of averaging over the probability measure \eqref{eq_total_prob}, since it involves handling simultaneously all the correlated variables over their past history, as standard in path-integral formulations. In fact, the idea is to replace the mean expectation of the internal field produced by the peer units and appearing in the exponent of the Boltzmann-Gibbs representation \eqref{eq_total_prob}, with its empirical evaluation. As we will show, this makes the measure $P_A(\bsi)$ factorized in such a way that we can recast the Lagrangian constraints in an approximated linear system that is feasible for a straightforward evaluation of the parameters $\bsJ$ and $\bsh$. 

Let us implement the plan. Suppose that we experimentally observe a system of $N$ particles at regular time
intervals for a total of $\maxtimes+1$ timepoints, then, according
to \cref{eq:vel_dir}, we can evaluate the velocity $\bsv$ and  the direction $\bsi$ for each particle and time point, overall collecting
the dataset:
\begin{equation}
\bsi^{\rm E}=\{\bsi_{i}^{{\rm E},t}\,|1 \leq i \leq N, \,\,1 \leq t \leq \maxtimes\},
\end{equation}
hence, we can calculate 
\begin{eqnarray}
&&\boldsymbol m_t^E \equiv \boldsymbol m^{\rm E}(\bsi^{{\rm E},t}) = \frac{1}{N} \sum_{i=1}^N \bsi_{i}^{{\rm E},t},\\
&&\boldsymbol M^E \equiv \boldsymbol M^{\rm E}(\boldsymbol s^{\rm E}) = \frac{1}{N_T - \tau_M} \tsum t \boldsymbol m^{\rm E}(\bsi^{{\rm E},t}),\\ \small
&& R^E (\tau)  = \frac{1}{\maxtimes - \tau_M}  \tsum t \boldsymbol m^{\rm E}(\bsi^{{\rm E},t}) \cdot \boldsymbol m^{\rm E}(\bsi^{{\rm E},t-\tau}), 
\end{eqnarray}
where the superscript E highlights that the quantity is evaluated by experimental data and in the last definition we wrote $R^E (\tau)\equiv R^{\rm E}(\tau, \boldsymbol s^{\rm E})$ to ligthen notation.
In order to estimate the parameters $\bsJ$ and $\bsh$ appearing in \cref{eq_total_prob},
and thus build a model for the dataset $\bsi^{\rm E}$, we can use the
method of maximum likelihood \cite{Cox2011}. The log-likelihood $l$ of the model  \eqref{eq_total_prob} given the dataset $\bsi^{\rm E}$ is
\begin{align}
& l(\bsJ,\bsh|\bsi^{\rm E})  \equiv   \log P(\bsi^{\rm E}|\bsJ,\bsh)\nn\\
 & =  \tsum t\sum_{i=1}^{N}\bsi_{i}^{{\rm E},t}\cdot\left(\frac{1}{N}\dsum{\tau}J_{\tau}\sum_{j=1}^{N}\bsi_{j}^{E,t-\tau}+\bsh\right)\label{eq:exact_loglikelihood}\\ \small
 &   -\log\int\mathcal{D}\bsi e^{[\tsum t\sum_{i=1}^{N}\bsi_{i}^{t}\cdot (\frac{1}{N}\dsum{\tau}J_{\tau}\sum_{j=1}^{N}\bsi_{j}^{t-\tau}+\bsh)]}.\nn
\end{align}
The values of $\bsJ$ and $\bsh$ that maximize $l(\bsJ,\bsh|\bsi^{\rm E})$ represent the maximum likelihood estimate for the parameters. We therefore derive $l(\bsJ,\bsh|\bsi^{\rm E})$ in \cref{eq:exact_loglikelihood} and solve for
\be \label{eq:ml}
\frac{\partial l}{\partial\bsJ}= \bm 0, \ \ \frac{\partial l}{\partial\bsh}=\bm 0.
\ee
However, such a direct approach requires the evaluation of high-dimensional multi-variable integrals, like the one appearing in the last term of \cref{eq:exact_loglikelihood},  which entail several numerical issues as for  the stability of the solution and the computational time (e.g., in our experimental datasets the number of correlated variables, for which we have to perform integrals, is $\mathcal {O}({10}^{5})$, being $\maxtimes = 150$ and $N =300$).
%
%
%
%
Thus, we resort to an approximation: First, in \cref{eq_total_prob} we replace the internal field, acting on $\bsi_i^t$, with its empirical counterpart
\be\label{eq2}
\frac{1}{N}\dsum{\tau}J_{\tau}\sum_{j=1}^{N}\bsi_{j}^{t-\tau} \rightarrow \dsum{\tau}J_{\tau}\bsm_{t-\tau}^{\rm E},
\ee
along the  lines of the mean-field approximation in statistical mechanics. 

With the substitution \eqref{eq2}, pairs like $\bsi_i^t \bsi_j^{t-\tau}$ no longer appear in \eqref{eq:exact_loglikelihood}, that is, in a statistical-mechanics jargon, the original two-body model has been recast into a one-body model. Also, with this passage, the dependence displayed by the probability distribution on the first tranche of trajectories, i.e., $\bsi_{i}^{t},\; t = 1,\cdots, \delays$ is lost, but it can be restored, at least formally, by introducing some constraints that set the missing variables to their experimental values, in such a way that $P_{\rm A}(\bsi)$ remains defined over the whole set of variables:
\begin{eqnarray}\nonumber
P(\bsi)  \rightarrow P_{\rm A}(\bsi) &\equiv& \frac{1}{Z_{\rm A}}\prod_{i=1}^{N}\left\{ \tprod t e^{[\bsi_{i}^{t}\cdot (\dsum{\tau}J_{\tau}\bsm_{t-\tau}^{\rm E}+\bsh )]}\right\} \\
&\cdot& \left[ \prod_{t=1}^{\delays}\delta\left(\bsi_{i}^{t}-\bsi_{i}^{{\rm E},t}\right)\right] ,\label{eq:prob_approx}
\end{eqnarray}
with
\begin{equation}\label{eq_za}\small
Z_{\rm A} \equiv \int \left( \prod_{t=\delays+1}^{\maxtimes}\prod_{i=1}^{N}d\bsi_{i}^{t} \right) \exp\left[\bsi_{i}^{t}\cdot\left(\dsum{\tau}J_{\tau}\bsm_{t-\tau}^{\rm E}+\bsh\right)\right],
\end{equation}
where the subscript $\rm A$ stands for \emph{approximated}.
In this way, we are still able to evaluate all the momenta of the variables $\boldsymbol s$, including those involving $\bsi_{i}^{t},\; t = 1,\cdots, \delays$.


We denote by $\langle\rangle_{\rm A}$ the average performed
with respect to $P_{\rm A}$, i.e., for the generic observable $F(\bsi)$,
\begin{equation}
\langle F(\bsi)\rangle_{\rm A}\equiv\int\mathcal{D}\bsi P_{\rm A}[\bsi]F(\bsi),
\end{equation}
and note that 
\begin{align}
\langle\bsm_{t}\rangle_{\rm A}&=\int \left( \prod_{t=\delays+1}^{\maxtimes}\prod_{i=1}^{N}d\bsi_{i}^{t} \right) P_{\rm A}(\bsi)\left(\frac{1}{N}\sum_{i=1}^{N}\bsi_{i}^{t}\right)=\nn\\
&=\begin{cases}
\bscalM\left(\dsum{\tau}J_{\tau}\bsm_{t-\tau}^{\rm E}+\bsh\right) & \delays\le t\le\maxtimes\\
\bsm_{t}^{\rm E} & 1\le t\le\delays
\end{cases}\label{eq:magn_average}
\end{align}
where the vector function $\bscalM$ is defined as
\begin{equation}
\bscalM(\bsx)\equiv\frac{\bsx}{\left| \bsx\right| }\frac{\mathcal{I}_{1}\left(\left| \bsx\right| \right)}{\mathcal{I}_{0}\left(\left| \bsx\right| \right)}
\end{equation}
 and $\mathcal{I}_{0},\mathcal{I}_{1}$ are hyperbolic Bessel functions (also known as modified Bessel function of the first kind) of order $0$ and $1$, respectively \cite{abramowitz1965handbook}. 
In addition, we note that $P_{\rm A}$ is factorized with respect to every variable $\bsi_{i}^{t}$, thus any average of products of magnetizations at
different times $t$ and $t^{\prime}$ equals a product
of averages:
\begin{equation}\label{factor_m}
\langle\bsm_{t}\bsm_{t^{\prime}}\rangle_{\rm A}=\langle\bsm_{t}\rangle_{\rm A}\langle\bsm_{t^{\prime}}\rangle_{\rm A}.
\end{equation}
The approximation \eqref{eq:prob_approx} and the property  \eqref{factor_m} allow us to rearrange \cref{entropy_problem2,entropy_problem3} in the following form:
\begin{align}
\label{cstr1}
&\tsumnorm t\langle\bsm_{t}\rangle_{\rm A}   =  \bM^{\rm E}=\tsumnorm t\bsm_{t}^{\rm E},\\
\label{cstr2}
&\tsumnorm t\langle\bsm_{t}\rangle_{\rm A}\cdot\langle\bsm_{t-\tau}\rangle_{\rm A}   =  \autocorr^{\rm E}(\tau)= \\ \nonumber
& \tsumnorm t\bsm_{t}^{\rm E}\cdot\bsm_{t-\tau}^{\rm E},~~  1 \le \tau \le \delays,
\end{align}
where we dropped the explicit dependence on $\bsi^{E,t}$ to lighten the notation.
These equations are simultaneously fulfilled by requiring that
\begin{equation}
\label{eq:suff}
\langle\bsm_{t}\rangle_{\rm A}=\bsm_{t}^{\rm E},\qquad  \delays + 1 \leq t \leq \maxtimes. 
\end{equation}
In fact, given that Eq.~\eqref{eq:suff} is automatically satisfied for $1 \leq t \leq \delays$, if  Eq.~\eqref{eq:suff} holds, then Eqs.~(\ref{cstr1})-(\ref{cstr2}) are satisfied too. 
%

We stress that  \eqref{eq:suff} is only a \emph{sufficient} condition for  \cref{cstr1,cstr2} to hold since this condition introduces stricter constraints than those stemming from~\cref{cstr1,cstr2}. However, by moving from \cref{cstr1,cstr2} to \cref{eq:suff} we retain the very same parameters, that are $\bsh$ and $\bsJ$. Specifically, in this problem, the amount of data available $\maxtimes \times N$ is typically much larger than the number of parameters $D + \delays$. Thus, this stricter condition does not imply any overfitting (we are introducing constraints and not parameters) which, on the other hand, could be controlled by tuning $\delays$.

Exploiting \eqref{eq:magn_average}, \Cref{eq:suff} can be rewritten as
\begin{equation}\label{eq_2}
\bscalM\left(\dsum{\tau}J_{\tau}\bsm_{t-\tau}^{\rm E}+\bsh\right)=\bsm_{t}^{\rm E},\qquad \delays +1 \leq t \leq \maxtimes.
\end{equation}
Since $\bscalM$ is a monotone function, its inverse function is well defined and we can rewrite \eqref{eq_2} as
\begin{align}
\bsG_{t}^{\rm E} & \equiv  \bscalM^{-1}\left(\bsm_{t}^{\rm E}\right),\\
\dsum{\tau}J_{\tau}\bsm_{t-\tau}^{\rm E}+\bsh & =  \bsG_{t}^{\rm E},\qquad \delays +1 \leq t \leq \maxtimes. \label{eq:linear_system}
\end{align}
We observe that the operator $\bsG_{t}^{\rm E}$ is  linear in the variables $\bsJ,\bsh$: this property turns out to be crucial for the following developments. Let us underline that \eqref{eq:linear_system} constitutes a linear system, whose solution provides us with the parameters $\bsJ,\bsh$, but this system involves more constraints  than parameters. More precisely, as a result of \eqref{eq:suff}, the number of parameters is still $D + \delays$, and the number of constraints is now $(\maxtimes - \delays) \times D$.  We can therefore enrich the model by stating that the relation \eqref{eq:linear_system} does not hold deterministically and, thus, introduce a source of noise represented by a standard Gaussian variable tuned by the additional parameter $\sigma$.
%
This stochastic term turns the linear system \eqref{eq:linear_system} into an auto-regressive model\footnote{In an autoregression model, we forecast the variable of interest using a linear combination of past values of the variable; the term autoregression indicates that it is a regression of the variable against itself, see e.g., \cite{Theodoridis-2020}.} of order $\delays$
\begin{align}
\dsum{\tau}J_{\tau}\bsm_{t-\tau}^{\rm E}+\bsh+\sigma\boldsymbol{\epsilon}_{t} & =  \bsG_{t}^{\rm E},\qquad \delays +1 \leq t \leq \maxtimes,\label{eq:linearised_model}\\
\boldsymbol{\epsilon}_{t} & \sim  \mathcal{N}\left(\boldsymbol{0}_{\dims},\boldsymbol{1}_{\dims}\right),\quad\sigma\ge0.
\end{align}
In other words, given the parameters $\bsJ,\bsh$ and $\sigma$, the magnetization $\bsm_{t}$ at time $t$ depends on the magnetization exhibited in the previous $\delays$ times according to the evolution rule:
\begin{equation}
\bsm_{t}=\bscalM\left[\bsh+\dsum{\tau}J_{\tau}\bsm_{t-\tau}+\sigma\boldsymbol{\epsilon}_{t}\right].\label{eq:fast_dynamics}
\end{equation}
Now, the log-likelihood $l_{\rm A}(\bsJ,\bsh,\sigma |\bsi^{\rm E})$ related to the auto-regressive model is introduced as
\begin{equation}\small
-\frac{1}{2\sigma^{2}}\tsum t\left| \dsum{\tau}J_{\tau}\bsm_{t-\tau}^{\rm E}+\bsh-\bsG_{t}^{\rm E}\right| ^{2}-\left(\avtimes\right)\frac{\dims}{2}\log\left(2\pi\sigma^{2}\right)\label{eq:gaussian_loglikelihood}
\end{equation}
and, in the next section, we will extremize $l_{\rm A}(\bsJ,\bsh,\sigma |\bsi^{\rm E})$ to obtain an estimate for $\bsJ, \bsh$, and $\sigma$.

\subsubsection{Analytical solution for the estimated parameters\label{subsec:solution}}

As customary in maximum likelihood estimations, we evaluate the gradient of the log-likelihood $l_{\rm A}(\bsJ,\bsh,\sigma |\bsi^{\rm E})$ and set it to zero
in order to estimate $\bsJ,\bsh$ and $\sigma$. To this goal, we first compute the derivatives of \eqref{eq:gaussian_loglikelihood}  
\begin{align}\small
\nabla_{\bsh}l_{\rm A} & =  -\frac{\avtimes}{\sigma^{2}}\left\langle \bsh+\dsum{\tau}J_{\tau}\bsm_{t-\tau}^{\rm E}-\bsG_{t}^{\rm E}\right\rangle _{t},\label{eq:grad_h}\\ \small
\nabla_{J_{\tau}}l_{\rm A} & =  -\frac{\avtimes}{\sigma^{2}}\left\langle \left(\bsh+\dsum{\lambda}J_{\lambda}\bsm_{t-\lambda}^{E}-\bsG_{t}^{\rm E}\right)\cdot\bsm_{t-\tau}^{\rm E}\right\rangle _{t},\label{eq:grad_j}\\ \small
\nabla_{\sigma}l_{\rm A} & =  \frac{\avtimes}{\sigma^{3}} \langle \left| \dsum{\tau}J_{\tau}\bsm_{t-\tau}^{\rm E}+\bsh-\bsG_{t}^{\rm E}\right| ^{2}\rangle _{t}-\left(\avtimes\right)\frac{\dims}{\sigma},\label{eq:grad_sigma}
\end{align}
where \begin{equation}
\tavg{ \cdot }\equiv\tsumnorm t \cdot
\end{equation}
denotes the time average. 
\newline
Setting \eqref{eq:grad_h} to zero and solving for $\bsh$, we get
\begin{equation}\label{H-inferred}
\bsh=\langle \bsG_{t}^{\rm E}\rangle _{t}-\dsum{\tau}J_{\tau}\left\langle \bsm_{t-\tau}^{\rm E}\right\rangle _{t}.
\end{equation}
Proceeding along the same lines, from \eqref{eq:grad_j}, for $\bsJ$ we obtain
\begin{equation}\label{LaSolution}
\dsum{\tau}A_{\lambda\tau}J_{\tau}=
B_{\tau}\leftrightarrow\bsJ\equiv\boldsymbol{A}^{-1}\cdot\boldsymbol{B},
\end{equation}
where the $\delays \times \delays$ matrix $A_{\lambda\tau}$ and the $\delays$-dimensional vector $B_{\lambda}$ are
\begin{align}
A_{\lambda\tau} & \equiv  \left\langle \bsm_{t-\tau}^{\rm E}\cdot\bsm_{t-\lambda}^{\rm E}\right\rangle _{t}-\left\langle \bsm_{t-\tau}^{\rm E}\right\rangle _{t}\cdot\left\langle \bsm_{t-\lambda}^{\rm E}\right\rangle _{t},\\
B_{\lambda} & \equiv  \langle \bsG_{t}^{\rm E}\cdot\bsm_{t-\lambda}^{\rm E} \rangle _{t}-\langle \bsG_{t}^{\rm E}\rangle _{t}\cdot \langle \bsm_{t-\lambda}^{\rm E} \rangle _{t}.
\end{align}
Finally, setting $\nabla_{\sigma}l_{\rm A}=0$, we obtain the variance
$\sigma^{2}$ 
 of the auto-regressive process \eqref{eq:fast_dynamics}:
\begin{equation}
\sigma^{2}=\frac{1}{\dims}\left\langle \left| \bsh+\dsum{\tau}J_{\tau}\bsm_{t-\tau}^{\rm E}-\bsG_{t}^{\rm E}\right| ^{2}\right\rangle _{t}.\label{eq:unexplained_variance}
\end{equation}

\subsubsection{Errors on the inferred measurements\label{subsec:error}}

In order to estimate the errors for the inferred parameters in the maximum likelihood estimation,
 we exploit the Fisher information matrix (see e.g., \cite{Frieden-2004}). 
 We recall that, given some dataset $\boldsymbol X$  where each observation $\bsx_{i}$ is assumed to be identically and independently distributed according to a true underlying distribution, and being $f_{\bstheta}(\boldsymbol x)$ a model probability density function parametrized by $\bstheta$, we can write the log-likelihood function as $l(\bstheta | \bsx) = \sum_{i=1}^n \log f_{\bstheta}(\boldsymbol x_{i})$, where $n$ is the sample size.
Then, the (empirical) Fisher information matrix $\boldsymbol I$ has elements given by
\begin{equation}
\label{eq:I_50}
{\boldsymbol I}_{\theta_a,\theta_b} \equiv -\int \exp[l(\bstheta | \bsx)] \frac{\partial^{2}l(\bstheta | \bsx)}{\partial \theta_{a}\partial\theta_{b}} d\bsx,
\end{equation}
namely its elements correspond to the expectation of the elements of the Hessian matrix of the log-likelihood. 
Let us suppose that the true parameter is $\bstheta_0$, and that the maximum-likelihood estimate of $\bstheta_0$ is ${\bstheta}^* = \textrm{argmax}_{\bstheta} l(\bstheta | \bsx)$.
Then, one can prove that ${\bstheta}^* \sim \mathcal N({\bstheta}_0, {\boldsymbol I}^{-1}({\bstheta}_0))$. Since when the sample size approaches infinity, the maximum-likelihood estimate approaches the true parameter (this is also known as the consistency property), we can write that the covariance matrix for the estimated parameters is just 
\begin{equation}
\boldsymbol{\Sigma}_{{\bstheta}^*} \underset{n \gg1}{\approx} {\boldsymbol I}^{-1}({\bstheta}^*).\label{eq:fisher_vs_errors}
\end{equation}
We will use this result to provide the error estimates for our model. By \eqref{eq:gaussian_loglikelihood} the
matrix elements of the Fisher information read
\begin{align}
I_{H_a,H_b} & =  \frac{\avtimes}{\sigma^{2}}\delta_{ab},\label{eq:fisherscore}\\
I_{J_{\tau},\bsh} & =  \frac{\avtimes}{\sigma^{2}}\left\langle \bsm_{t-\tau}^{\rm E}\right\rangle _{t},\\
I_{J_{\tau},J_{\lambda}} & =  \frac{\avtimes}{\sigma^{2}}\left\langle \bsm_{t-\tau}^{\rm E}\cdot\bsm_{t-\lambda}^{E}\right\rangle _{t},\\
I_{\sigma,\sigma} & =  \frac{\avtimes}{\sigma^{2}}~2\dims,\\
I_{\sigma,J_{\tau}} & =  I_{\sigma,\bsh}=0, 
\end{align}
 and \eqref{eq:I_50} becomes
\begin{equation}\label{eq3}
\boldsymbol{I} = \frac{\avtimes}{\sigma^{2}}\left(\begin{array}{ccc}
\left\langle \bsm_{t-\tau}^{\rm E}\cdot\bsm_{t-\lambda}^{E}\right\rangle _{t} & \left\langle \bsm_{t-\tau}^{\rm E}\right\rangle _{t} & \boldsymbol 0\\
\left\langle \bsm_{t-\lambda}^{E}\right\rangle _{t} & {\boldsymbol \delta} & {\boldsymbol 0}\\
\boldsymbol 0 & {\boldsymbol 0} & 2\dims
\end{array}\right).
\end{equation}
Finally, by substituting \cref{eq3} into \cref{eq:fisher_vs_errors}, we obtain the
error estimates for $\bsJ,\bsh$ and $\sigma$:
\begin{equation}
\textrm{Var}\left(\bsJ,\bsh,\sigma\right)=\textrm{diag}\left(\boldsymbol{\Sigma}_{{\bstheta}^*}\right).
\end{equation}
In the section dedicated to the model validation, we  use this relation to test the goodness of our method on synthetic datasets.

\subsubsection{Effective number of delayed interactions\label{sec:delays}}

So far, in our analysis, we assumed that the parameter $\delays$ is a known constant. In most practical inference scenarios, where one needs to find  from scratch the best model for a given dataset, the number of the effective delayed interactions $\delays$  that must be considered is unknown:  in what follows, we will provide a way to estimate it.
\newline
In order to tackle this problem, we rely on the Akaike Information Criterion (AIC)  \cite{aic_banks2017}, which is used in the literature to probe the flexibility of an auto-regressive model. 
In fact, given a dataset and a model with a mean squared error $\epsilon^{2}$, $n_{\rm p}$ parameters and $n_{\rm o}$ observations, the quantity
\begin{equation}
\akaike\equiv\frac{2n_{\rm p}}{n_{\rm o}}+\log\left(\epsilon^{2}\right)\label{eq:generic_akaike}
\end{equation}
estimates the prediction error of the model. Thus, having developed several models for a certain dataset, the AIC estimates the quality of each model relative to all others, hence providing a tool for model selection.  Note that, by definition, the AIC favours those models yielding a small error \emph{and} a small number of parameters.

In the analysis above, the parameters appearing in \cref{eq:generic_akaike} are
\begin{equation}
n_{\rm o}  =  \dims\left(\maxtimes-\delays\right),~~~~ n_{\rm p}  =  \delays+\dims+1,~~~~ \epsilon^{2}  =  \frac{\maxtimes}{\maxtimes-\delays}\sigma^{2}
\end{equation}
Thus,
\begin{equation}
\akaike=2\frac{\delays+\dims+1}{D\left(\avtimes\right)}+\log\left(\frac{\maxtimes ~ \sigma^{2}}{\maxtimes-\delays}\right).\label{eq:taylored-akaike}
\end{equation}
To summarize, the protocol followed in the next sections is the following: we will perform inference for multiple values of $\delays$ and we solve for the fields and the couplings via eq.s \eqref{H-inferred}\eqref{LaSolution}. We further determine $\sigma^2$ from \cref{eq:unexplained_variance}, next, we will evaluate \eqref{eq:taylored-akaike} for each of the inferred models, and we will select the value of $\tau_M$ that implies the smallest estimate of $\akaike$. Finally, the maximal $J_{\tau}$ in the selected time window $\tau \in [0, \tau_M]$ turns out to be the best estimate (inferred value) of the actual coupling (planted value).

\section{Model validation\label{sec_val}}

In the remaining of this Appendix, we test our inference method on synthetic and real datasets: we consider two synthetic datasets, generated from the HKM and from the VM, and a biological dataset, collected via time-lapse microscopy on dendritic cell migration in a chemoattractant field (see \cite{Alemanno2023} for details).   
\newline
In general, we first run the inference protocol on such datasets and obtain the parameters $\bm{J}$ and $\bm{H}$, which we call \textit{inferred} parameters, and then we evaluate the goodness of the inference process.  As for the HKM and the VM, previously addressed, a validation is possibile by comparing the inferred parameters and those set in the model under consideration and denoted as \textit{planted} parameters. The HKM is the direct formulation of the inverse problem developed in this paper, thus, in this setting, \textit{inferred} and \textit{planted}  parameters  must match even quantitatively (and their agreement is a {\em conditio sine qua non} to move to more challenging scenarios); in fact,  in this case, inference is excellent. Conversely, for the VM, the inference process can be harder as this model is not the direct model of the delayed ME approach we developed, and, in order to investigate the problem meticulously, we face two versions of the delayed VM, namely, the topological and the metric ones. In both cases we have a set of units whose dynamics is non-Markovian and influenced by the trajectories of a subset of peers, made of $i$. the closest $n_c$ units (regardless of their reciprocal distance) in the former version, or $ii$. the closest units within a certain interaction range $r$  (regardless of their number) in the latter version.
\newline
Finally, the analysis of the biological datasets aims to be a proof of concept to highlight the non-Markovianity of biological dynamics: this is why we consider for the test study the migration of dendritic cells as the latter are expected to perceive signalling proteins that must diffuse from the emitter cell to the receiver one, thus, these interactions are natural prototypes of delayed interactions. 

We anticipate that, beyond successfully testing our delayed-inference protocol on all the synthetic datasets, we obtained a remarkable result in inspecting the biological ones: the trajectory of a given dendritic cell at time $t$ turns out to be influenced by those of the other cells for a past temporal window spanning back up to $\tau_M \sim \mathcal O(10)$ time-steps, thus resulting in severe violation of the Markovian assumption.

\subsection{Synthetic dataset: the Heisenberg-Kuramoto model\label{sec:me_vs_me}}

The HKM with delayed interactions can be defined by interpreting the probability distribution \eqref{eq_total_prob}, with normalization given by \eqref{NormaliZZa}, as the Gibbs measure $\exp(-\mathcal{H}({\boldsymbol s | \bsJ, \bsh}))$  of the Hamiltonian
\begin{equation}\label{HamHK}
\mathcal{H}({\boldsymbol s |  \bsJ, \bsh})= \sum_{i=1}^{N}\bsi_{i}^{t}\cdot\left(\frac{1}{N}\dsum{\tau}J_{\tau}\sum_{j=1}^{N}\bsi_{j}^{t-\tau}+\bsh\right),
\end{equation}
where the dynamical variables are the spins $\boldsymbol s$, while the parameters $\bsJ$ and $\bsh$ represent, respectively, the interactions perceived with a delay $\tau$ (back in the history up to $\tau = \tau_M$) and the external field. 

To setup the validation of our inferential protocol  on the HKM we perform the following steps: 
\begin{enumerate}
\item We set $D=2$ and $\delays = 4$, and draw the planted parameters $J_{1},\cdots,J_{\delays},H_{x},H_{y}$ independently
and identically from $U(-1,1)$, where $U(a,b)$ denotes the uniform probability distribution between $a$ and $b$. 
\item With the parameters just set, we sample the trajectories from the Boltzmann-Gibbs distribution related to the Hamiltonian \eqref{HamHK}
from $t=0$ to $t= 5 \times 10^3$, and store the magnetizations from $t= 1 \times 10^3$
to $t=5 \times 10^3$  in order to consider equilibrated snapshots. 
\begin{enumerate}
\item  the simplest (despite rather lenghty) way of generating trajectories is to create the configuration  $\bsi^{t}$ by collecting all the vectors $\bsi^{t-1},...,\bsi^{t-\tau_M}$ and inserting them into the Hamiltonian $\eqref{HamHK}$. Then we generate, trought its Boltzmann-Gibbs weight, the configuration        
$\bsi^{t}$ and so on for  $\bsi^{t+1}$
\item however, as we just need their averages (i.e. $\bsm_{t}$ as a function of $\bsm_{t-1}, ..., \bsm_{t-\tau_M}$), an alternative (faster) route consists in using the coarse-grained evolution coded by \eqref{eq:fast_dynamics} and iterate the latter (that is equivalent to the previous one as the model is mean-field).
\end{enumerate}
Note that trajectories are not sampled  {\em as a whole}, rather gradually built up time step by time step, such that their generation is not a  unique high-dimensional sample from $P(\boldsymbol{s})$, rather a stochastic process with memory. 
\item We use the stored magnetizations to carry out the inference process
according to our formulation (see eq. \eqref{H-inferred} for the fields, eq. \eqref{LaSolution} for the couplings and  eq.  \eqref{eq:unexplained_variance} for the variance) , obtaining the estimates of $,H_{x},H_{y},J_{1},\cdots,J_{\delays},\sigma$.  $\delays$ is found by using the Akaike information criterion (see eq. \eqref{eq:taylored-akaike} and its related discussion).
\item We store the corresponding pairs of \textit{inferred} and \textit{planted} parameters. 
\end{enumerate}
This process is repeated $150$ times and the collected data are summarized in Fig.~\ref{fig:me_vs_itself}. In particular, \cref{fig:me_vs_itself}a shows that 
 the planted and inferred parameters are arranged along the diagonal $x=y$, implying a quantitatively excellent estimate for parameters. 
In addition, \cref{fig:me_vs_itself}d shows that the inference framework is able to exactly detect the extent of the time window $[0, \tau_M]$ most of the times; although in some cases this is slightly overestimated, we verified that the magnitude of the extra interactions $J_{\tau > \tau^{\rm planted}}^{\rm inferred}$ is statistically compatible with zero within the error on the planted data.
This can be seen from Fig.~\ref{fig:me_vs_itself}b: as $J_{\tau > \tau^{\rm planted}}^{\rm planted} \equiv 0$, if $J_{\tau > \tau^{\rm planted}}^{\rm inferred} \neq 0$, its magnitude corresponds to the absolute error and in these cases we reveal only small errors ($\sim \mathcal O(10^{-2})$).

\subsection{Synthetic dataset: the Vicsek model\label{sec:val_vicsek0}}

\subsubsection{The topological Vicsek model with delay\label{sec:vicsek1}}

\begin{figure} 
\begin{centering}
\includegraphics[width=0.48\textwidth]{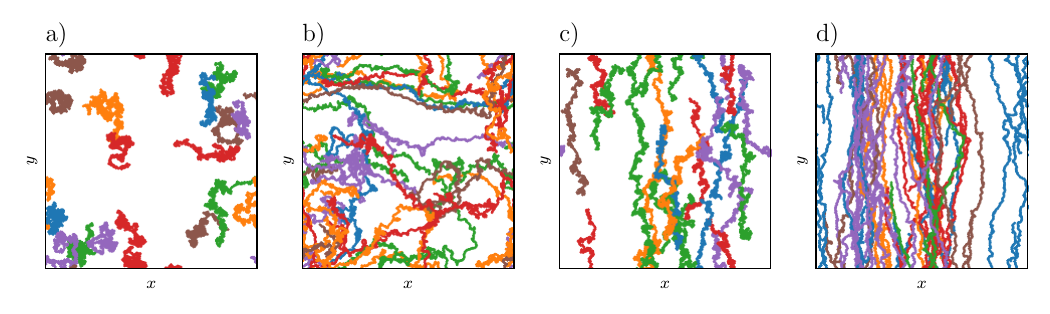} 
\par\end{centering}
\caption{\label{fig:Trajectories-Vicsec} Examples of trajectories generated by the VM. Simulations were implemented by letting $N=100$ particles free to move in a 2D square of linear size $L=150$ with periodic boundary conditions. Further, we set $\delta t=1$, $\Delta = 6$ and $n_c=4$. Panels: a)  Trajectories sampled at $J=0, H_x=H_y=0$ (as in a standard random walk). b)   Trajectories sampled at $J>0, H_x=H_y=0$ (as in a correlated random walk). c)   Trajectories sampled at $J=0,H_x=0, H_y>0 $ (as in a drifted random walk). d)  Trajectories sampled at $J>0$ and $H_x=0$, $H_y>0$ (as in  a correlated and drifted random walk).}
\end{figure}

In this section, we consider the topological version of the  VM \cite{vicsek1995novel,ginelli2010relevance}, equipped with a delay $\Delta$ in signalling among different units. Such a model is defined by the following dynamical equations 
\begin{eqnarray}\label{eq:vicsek_dyn1}   
D_{ij}&=&(\Delta-1) (1-\delta_{i}^{j})\\
\bsv_{k}^{t+1}&=&v_{0}\boldsymbol{\Gamma}\Bigg(\!\frac{J}{\left|n_{k}^{t}\right|}\sum_{j\in n_{k}^{t}}\!\!\frac{\bsv_{j}^{t-D_{ik}}}{v_{0}}+\!\bsh\!+\!\sqrt{\delta t}\,\boldsymbol{\eta}_{k}^{t}\!\Bigg),\\  
\label{eq:vicsek_dyn2}
\bsx_{k}^{t+1}&=&\bsx_{k}^{t}+\delta{t}\,\bsv_{k}^{t+1} ~~~k=1,\cdots,N.
\end{eqnarray}
where $\bsv_{i}^{t}$, $\bsx_{i}^{t}$  are the velocity and the position of the $i$th particle at time $t$, respectively. 

Note that $n_{\rm c}$ and $\{n_{1}^{t}, \cdots, n_{n_{\rm c}}^{t} \}$  are, respectively, the  total number of topological neighbors and the set of topological neighbors of the $i$th particle, 
 including the $i$th particle, at time $t$, and that, by selecting $\Delta =1$, the Markovian limit of the standard VM is recovered.  The normalization operator $\boldsymbol{\Gamma}$
\begin{equation}
\boldsymbol{\Gamma}(\bsv) \equiv \frac{\bsv}{\left| \bsv\right| }, 
\end{equation} 
ensures that the magnitude of the velocity of each unit is kept constant. Further, 
 $\bsh$ is the external field applied to each particle, and  $\boldsymbol{\eta}_{k}^{t}\sim\mathcal{N}(\boldsymbol{0},\boldsymbol{1})$ is an i.i.d. noise. In this model, the particles are self-propelled, i.e.,  $\left| \boldsymbol{v}_{j}^{t}\right| = v_{0}$. Note that, as discussed above, we denoted the planted parameter $\bsh$ with the same symbol as the magnetic field in the ME  distribution \eqref{eq1}, but they do not need to coincide. 
Finally, the magnetization of the VM is given by the expression 
\begin{equation}\label{M_VTN}
\bsm_t=\frac{1}{N v_0}\sum_{k=1}^{N}\bsv_{k}^{t},
\end{equation}
which is analogous to \cref{eq_m}. 


Examples of trajectories generated by this VM are shown in Fig. \ref{fig:Trajectories-Vicsec}.


The validation procedure for the ME inference in the case of the VM is different from that of the HKM:  in the case of the VM we have $J, H_x$ as free parameters. Despite  these parameters play the same role of the coupling and the field in the HKM, the way they enter in the update rule of the units \eqref{eq:vicsek_dyn1}-\eqref{eq:vicsek_dyn2} is  different from that of the ME model:  for the VM we lack a Hamiltonian formulation and direct and inverse problems are not {\em diagonal}, therefore a straight comparison between planted and inferred parameters is not expected to hold sharply. Nevertheless, the estimate of $J_{\textrm{inferred}}$ and $H_{\textrm{inferred}}$ is still very informative about the real values of the delayed couplings. In fact, as shown in Fig.~\ref{fig:Testing-VicsecTopA}$c$, for $J_{\textrm{planted}}=0$ the algorithm correctly returns $J_{\textrm{inferred}}=0$, then, for values of $J_{\textrm{planted}}>0$ ($J_{\textrm{planted}}<0$) the inference protocol returns positive (negative) values of $J_{\textrm{inferred}}$; we also notice slight overestimates in the positive branch as the external field gets larger and larger and this should be ascribed to an intrinsic limitation of the present protocol.  
\newline
As a sideline, we also stress that, since in our mean-field framework all units contribute equally, it is possible to obtain an estimate of the number of nearest neighbors $n_c$. Indeed, by a glance at Fig. \ref{fig:Testing-VicsecTopA}$b$ we see that there are mainly two contributions to the coupling, namely $J_{(\tau=1)}$ and $J_{(\tau=\Delta)}$: the former is due to the inertia of the unit mass and this contribution is intrinsically Markovian, while the latter is due to the contribution of all the other units within $n_c$, see Eqs.~\eqref{eq:vicsek_dyn1}-\eqref{eq:vicsek_dyn2}. Hence, we propose the formula $n_c \approx 1 +   J_{\tau=\Delta}/J_{\tau=1}$ to estimate the number of nearest neighbors and this looks a reasonable one, see Fig. \ref{fig:Testing-VicsecTopA}$d$.
\newline
Beyond these evidences, a standard validation method in these non-diagonal problems is to compare the lowest-order moments of some observables evaluated from the original trajectories generated via the VM with those sampled from the ME once the optimal values for $\bsJ$ and $\bsh$ are learnt. In particular,  we measure the magnetization $\bsM$ and the correlation function $R(\tau=\Delta)$ in both models, and check if their values are statistically compatible. Indeed, they are as shown in Fig.~\ref{fig:Testing-VicsecTopB}.
\newline
Furthermore, we also inspect the Markovian limit, namely the standard topological VM, where we set $\Delta=1$ in \eqref{eq:vicsek_dyn1}. In this case the inference protocol correctly returns that the trajectories are Markovian: we expect that only $J_{(\tau=1)}$ is significantly different from zero among the inferred couplings $J_{\tau}$ for all $\tau \in (1,...,\tau_M)$ of the delayed ME protocol: this prediction is confirmed too (data not shown).


To summarize, in general, the validation procedure is carried out by means of the following algorithm:
\begin{enumerate}
\item Draw  $H_x$ from $U(0,2)$ and $J$ from $U(-2,+2)$.
\item \label{item2} Run the dynamics \eqref{eq:vicsek_dyn1}-\eqref{eq:vicsek_dyn2} from $t=0$ to $t=2 \times 10^3$, storing the magnetization at each step (excluding the first $10^3$ that are used for equilibration).
\item \label{item3} Perform the ME inference process, using the magnetizations defined in \eqref{M_VTN} as input and infer the couplings $\boldsymbol J$ and the field $\boldsymbol H$.
\item Setting the parameters according to the previous step,  simulate the ME model from $t=0$ to $t=2 \times 10^3$, collecting the magnetization at each step (excluding the first $10^3$ that are used for equilibration).
\item For both frameworks, evaluate the  moments $\bsM$ and $\autocorr(\tau)$, and calculate their relative difference.
\end{enumerate}

\begin{figure}[tb]
\begin{centering}
\includegraphics[width=0.48\textwidth]{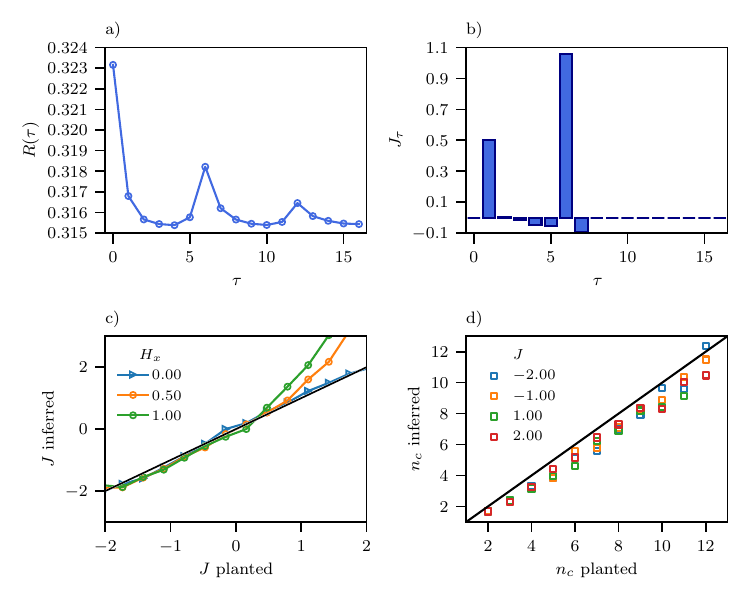} 
\par\end{centering}
\caption{\label{fig:Testing-VicsecMecA} Validation of our inference method with the metric Vicsek model (VM). The parameter setting is the same as in Fig.~ \ref{fig:Testing-VicsecTopA}. 
a) Example of the empirical correlation $R(\tau)$ vs $\tau$. b) Example of the (related) inferred coupling: note the presence of two peaks as in the topological case. c) Scatter plot of the inferred vs the planted values of the coupling, for several values of the (external) field $H_x$, as explained in the legend. d) Scatter plot of the nearest neighbors $n_c$ where the inferred estimate is obtained as $n_c = 1 + \frac{J_{\tau=\Delta}}{J_{\tau=1}}$.}
\end{figure}

\begin{figure}[tb]
\begin{centering}
\includegraphics[width=0.48\textwidth]{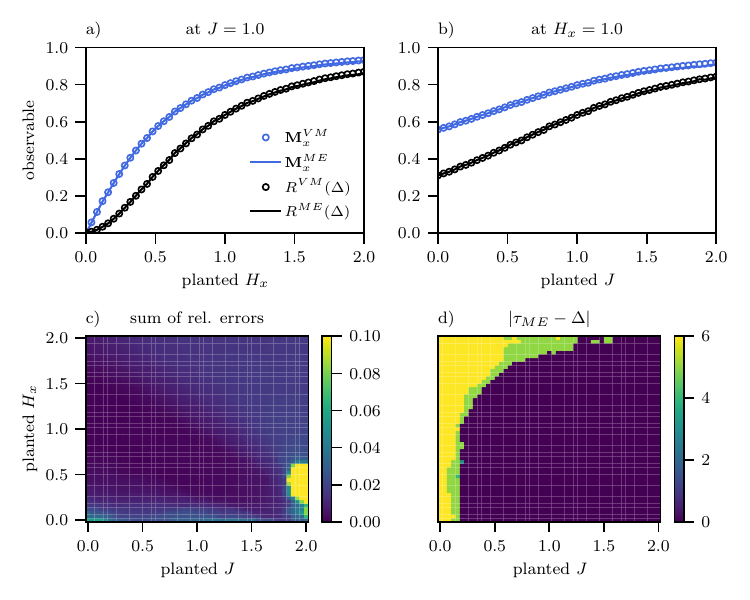} 
\par\end{centering}
\caption{\label{fig:Testing-VicsecMecB} Validation of our inference method with the metric Vicsek model (VM). The parameter setting is the same as in Fig.~ \ref{fig:Testing-VicsecTopA}.  
The superscripts $V$ and $ME$ pertain to quantities calculated for the VM or inferred via the ME method, respectively. a) Observables $\bsM$ and $R(\Delta)$ as functions of $H_x$, for the both models, with $J = 1.0$. b) Observables $\bsM$ and $R(\Delta)$ as functions of $J$, for both  models, with $H_x = 1.0$.
c) Sum of the relative errors $\frac{|M_{ME}-M_{V}|}{|M_{V}|}+\frac{|R_{V}(\Delta)-R_{ME}(\Delta)|}{R_{V}(\Delta)}$ as functions of $H_x$ and $J$, 
d) Absolute deviation between $\tau_{ME} := \text{argmax}_{\tau}|J_{ME}(\tau)|$ and the planted parameter $\Delta$ as a function of $J$ and $H_x$.
}
\end{figure}
\subsubsection{The metric Vicsek model with delay\label{sec:vicsek2}}

The solely difference between the metric version of the VM and its topological counterpart is that, for the former, the peers that contribute to the internal field for the velocity update of the $k$th unit (see \eqref{eq:vicsek_dyn1}-\eqref{eq:vicsek_dyn2}) are those units within a given radius and we fix  the interaction range to $r_{n_{c}} := \sqrt{\frac{n_{c}L^{2}}{\pi N}}$. The interaction range  has been defined as a function of $n_{c}$ in order to obtain comparable results for the two implementations. 
\newline
Apart from this criterion for neighbor selection, the dynamical evolution is the same as in the previous case. 
\newline
As in the topological counterpart, we perform inference of both coupling and field. Results are collected in Fig. \ref{fig:Testing-VicsecMecA} and a comparison of the lowest order statistics, e.g. $\boldsymbol{M}_x^{VM}$  vs $\boldsymbol{M}_x^{ME}$ and $R^{VM}$ vs  $R^{ME}$, is presented in Fig. \ref{fig:Testing-VicsecMecB}. As in the topological case, we can also inspect the number of units contributing to the internal field by the formula $n_c \sim 1 +   J_{(\tau=\Delta)}/J_{(\tau=1)}$, yet, in this metric case, there is a (very mild) systematic underestimate of $n_c$ beyond the over-estimate of the coupling in the presence of large field, as in the previous case (see panels $c$ and $d$ in Fig.~\ref{fig:Testing-VicsecMecA}).
Finally, mirroring the previous scenario, we also inspected in this metric situation if the Markovian limit (i.e. $\Delta=1$ in Eqs.~\eqref{eq:vicsek_dyn1}-\eqref{eq:vicsek_dyn2}) is captured by our ME dalayed inference protocol and this test turns out again to be successful as the algorithm return $J_{(\tau=1)}\neq 0$ only, as it should (data not shown).


\begin{figure}[H]
\begin{centering}
\includegraphics[width=0.5\textwidth]{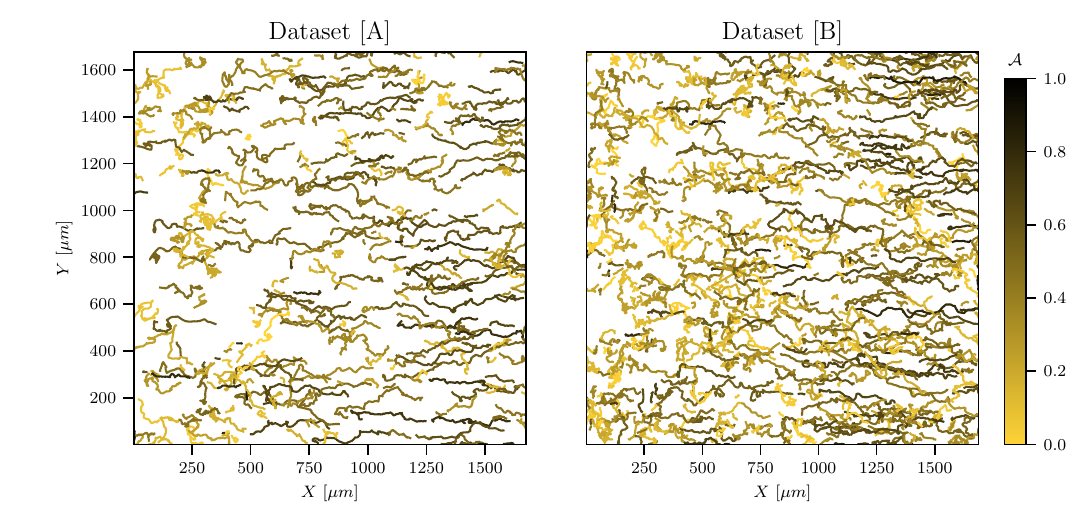}
\par\end{centering}
\caption{\label{fig:Real-Data-1} Trajectories of each cell through the whole observation time. For each trajectory a parameter that we call anisotropy $\mathscr{A}=2\frac{\max(|\Delta X_{\max}|,|\Delta Y_{\max}|)}{|\Delta X_{\max}|+|\Delta Y_{\max}|}-1\in[0,1]$
has been measured and used to paint tracks with the most anisotropy in black, and the least anisotropic tracks in yellow.}
\end{figure}

\subsection{Real dataset: dendritic cells in a chemokine gradient}\label{sec:dendro}

Having successfully tested our delayed ME inference on heterogeneous synthetic datasets,  we now apply it to an experimental dataset on cell migration  as a proof of concept. To be specific, in this section we  consider a two-dimensional tracking experiment where leukocytes move in the presence of chemokines (i.e. signalling proteins for white cells), which guide cell migration by  acting as a chemoattractant, see Fig.~\ref{fig:Real-Data-1} (the higher values of the gradient are in the right side of the pictures that is where cells are migrating). Some chemokines control cells of the immune system during immune-surveillance processes: for example, they may direct lymphocytes to lymph nodes so that they can control the invasion of pathogens, by interacting with antigen-presenting cells residing in these tissues: we consider this phenomenology as the most suitable to be investigated with this new inferential methodology. This is because, from one side, there is a vast network of signalling proteins (released by other cells) affecting immune dynamics \cite{Abbas}, in such a way that these leukocytes have to integrate all these diffusing signals to decide the next direction of motion and this may result in a delayed interaction because the diffusion of these proteins is slow if compared for instance with electric signals that are instead used by neurons to communicate or the electromagnetic field, the light, used by flocks. From the other side, there is a vast literature that implements Markov processes as a natural starting point to model immunodynamics, see e.g. \cite{NoMarkov1,NoMarkov2,NoMarkov3,Agliari-SciRep2014,Biselli-SciRep2017,agliari2020a}, thus confirming -- or otherwise disproving -- that their motion can be effectively described by a Markov chain is an important question in theoretical immunology. 


\begin{figure}[htbp]
\centering{}\includegraphics[width=0.5\textwidth]{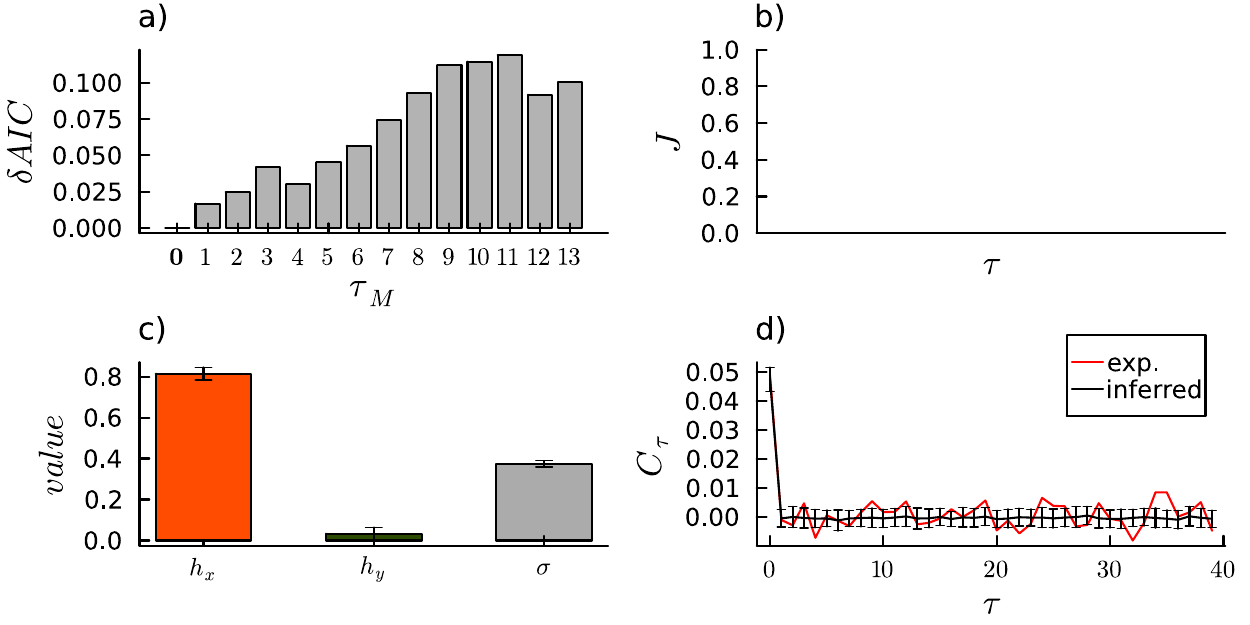}\\ \includegraphics[width=0.5\textwidth]{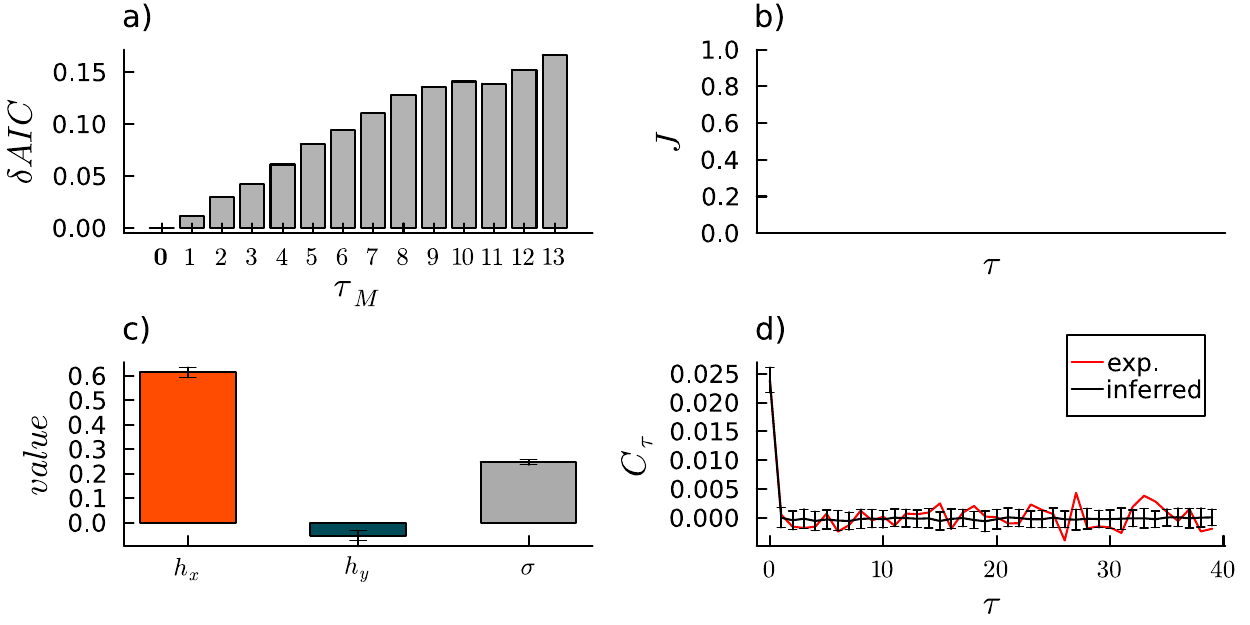}
\caption{\label{fig:NOinference} Test on the validity  of the delayed inference  procedure: we reshuffled trajectories in the chemokine-gradient dataset, for both regions A and B, and gave this random dataset back to the inferential protocol to inspect its outcomes. Panels: a) Difference between the AIC and its minimal value,  as a function of the total number of interactions $\delays$: note that this time the optimal $\tau_M$ is zero, in agreement with a Bernoullian process.
b) Inferred  values of the delayed interaction $J_{\tau}$ and  corresponding errors: note that, as expected, in this case there is no inferred coupling.
c) Inferred values for the two  components of the field, $H_x$ and $H_y$, 
and inferred value of $\sigma$. d) Correlation function $C(\tau)$, see \cref{eq:conn_corr},  from  the experimental data and from the  ME model. As expected, while permutation invariance over the frames destroyed the temporal correlation of the motion (induced by the couplings) it did not erase the gradient of the chemoattractant, that is still well inferred.}
\end{figure}

The system under investigation has been described in \cite{agliari2020a}: we collected  ordered time series of the positions of dendritic cells migrating via chemotaxis toward a chemoattractant (i.e. dendritic cells perceive, and follow, a citokine gradient released by the chemoattractant). Data was collected for  regions $\rm A$ and $\rm B$,  thus yielding two distinct datasets, see Fig.~\ref{fig:Real-Data-1}.

We apply the inferential ME framework to datasets A and B separately.
For the inferred ME model, we will show  
$$
\delta\akaike(\delays) \equiv \akaike(\delays)-\min_{\delays'}\akaike(\delays'),
$$ 
i.e., the difference between  $\akaike$ for a given $\delays$ and  $\akaike$ for the optimal value of $\delays$, as well as the inferred parameters $J_{\tau}$, $H_x, H_y$ and 
$\sigma$ and  the correlation function 
\begin{equation}\label{eq:conn_corr}
C(\tau) \equiv \tavg{\bsm(\bsi^{t})\cdot\bsm(\bsi^{t-\tau})} - \tavg{\bsm(\bsi^{t})}\cdot\tavg{\bsm(\bsi^{t-\tau})},
\end{equation}
with $1 \leq \tau \leq \delays$,  for the optimal value of $\delays$.
These results are shown in Fig.~\ref{fig:inference}  for both datasets A and B.
\newline
The results obtained for the two datasets are consistent and the unique picture that emerges can be streamlined as follows: the dynamics of a given leukocyte at a given time is influenced by the past actions of its peers for a very long time window (in these cases $\tau_M \sim 13,\ 14$), hence the trajectories that these cells paint while migrating are far from the Markovian limit. Moreover, along the dynamics, the inferred coupling are (mainly) negative in the recent past ($\tau=1,\ 2$) and (essentially) positive in the further past ($\tau \geq 3$) and the AIC criterion selects $\tau_M \sim 13,\ 14$. We speculate that a cell, during migration, may integrate signals it receives in time and, from this perspective, the present research seems to suggest that -- at least in the present setting -- dendritic cells may integrate signals for a temporal time-window of $\mathcal O(10)$  time steps, but clearly a dedicated study has to deepen this aspect in a forthcoming paper. 

Finally, as an additional test on the inference protocol on these real datasets, we also kept the collected ordered magnetizations  of the various frames and we reshuffled them to produce two new fake datasets (one per region) that we used to test the temporal-delayed protocol: results, reported in Fig. \ref{fig:NOinference}, correctly return the same external field (as that is a one-point correlation information that is not destroyed by the permutation of frames), while it selects as the optimal delay $\tau_{\textrm M}=0$ (that is correct as shuffling rules out any temporal correlation) and, coherently, a null value for the coupling.

\bibliographystyle{unsrt}

\begin{thebibliography}{10}

\bibitem{Locomotion2018}
A.~Biewener and S.~Patek.
\newblock {\em Animal locomotion}.
\newblock Oxford University Press, 2018.

\bibitem{Meijering-2012}
E.~Meijering, O.~Dzyubachyk, and I.~Smal.
\newblock Methods for cell and particle tracking.
\newblock {\em Methods in enzymology}, 504:183--200, 2012.

\bibitem{Procaccini2011}
A.~Procaccini et~al.
\newblock Understanding cell fate control by continuous single-cell
  quantification.
\newblock {\em Animal behaviour}, 82:759 --765, 2011.

\bibitem{Alemanno2023}
F.~Alemanno et~al.
\newblock Quantifying heterogeneity to drug response in cancer--stroma
  kinetics.
\newblock {\em Proc. Natl. Acad. Sci. USA}, page e2122352120, 2023.

\bibitem{jaynes1957information}
E.~T. Jaynes.
\newblock Information theory and statistical mechanics.
\newblock {\em Phys. Rev.}, 106(4):620, 1957.

\bibitem{seno2008maximum}
F.~Seno, A.~Trovato, J.~R. Banavar, and A.~Maritan.
\newblock Maximum entropy approach for deducing amino acid interactions in
  proteins.
\newblock {\em Phys. Rev. Lett.}, 100(7):078102, 2008.

\bibitem{lezon2006using}
T.~R. Lezon et~al.
\newblock Using the principle of entropy maximization to infer genetic
  interaction networks from gene expression patterns.
\newblock {\em Proc. Natl. Acad. Sci. USA}, 103(50):19033, 2006.

\bibitem{bialek2012statistical}
W.~Bialek et~al.
\newblock Statistical mechanics for natural flocks of birds.
\newblock {\em Proc. Natl. Acad. Sci. USA}, 109(13):4786, 2012.

\bibitem{aic_banks2017}
H.~T. Banks and M.~L. Joyner.
\newblock {AIC} under the framework of least squares estimation.
\newblock {\em Appl. Math. Lett.}, 74:33--45, 2017.

\bibitem{agliari2020a}
E.~Agliari et~al.
\newblock A statistical inference approach to reconstruct intercellular
  interactions in cell migration experiments.
\newblock {\em Science advances}, 6(11):eaay2103, 2020.

\bibitem{cavagna2014dynamical}
A.~Cavagna et~al.
\newblock Dynamical maximum entropy approach to flocking.
\newblock {\em Phys. Rev. E}, 89(4):042707, 2014.

\bibitem{abramowitz1965handbook}
M.~Abramowitz and I.~A. Stegun, editors.
\newblock {\em Handbook of mathematical functions: with formulas, graphs, and
  mathematical tables}.
\newblock National Bureau of Standards, 1972.

\bibitem{Frieden-2004}
B.~R. Frieden.
\newblock {\em Science from Fisher Information: A Unification}.
\newblock Cambridge Univ. Press., 2004.

\bibitem{VicsekDual}
L.~Barberis and V.E. Alban.
\newblock Evidence of a robust universality class in the critical behavior of
  self-propelled agents: Metric versus topological interactions.
\newblock {\em Phys. Rev. E}, (89(1)):012139, 2014.

\bibitem{vicsek1995novel}
T.~Vicsek et~al.
\newblock Novel type of phase transition in a system of self-driven particles.
\newblock {\em Phys. Rev. Lett.}, 75(6):1226, 1995.

\bibitem{DelayedVicsek}
D.~Geibb, K.~Kroy, and V.~Holubec.
\newblock Signal propagation and linear response in the delay vicsek model.
\newblock {\em Phys. Rev. E}, 5(106):054612, 2022.

\bibitem{ginelli2010relevance}
F.~Ginelli and H.~Chat{\'e}.
\newblock Relevance of metric-free interactions in flocking phenomena.
\newblock {\em Phys. Rev. Lett.}, 105(16):168103, 2010.

\bibitem{viscek+al_95}
T.~Vicsek and et al.
\newblock{Novel type of phase transition in a system of self-driven particles}
\newblock{\em Phys. Rev. Lett.}, 75.6, 1226, 1995. 

\bibitem{Cox2011}
D.R.~Cox 
\newblock{Principles of Statistical Inference}
\newblock{\em Cambridge University Press}, 2011.

\bibitem{Theodoridis-2020}
\newblock{Machine Learning, A Bayesian and Optimization Perspective}
\newblock{\em Academic Press}, 2020.


\bibitem{NoMarkov1}
K.~Althoff and J.~Degerman and C.~Wahlby and T.~Thorlin and J.~Faijerson and P.S.~Eriksson and T.~Gustavsson
\newblock{Time-lapse microscopy and classification of in vitro cell migration using hidden markov modeling}
\newblock{\em IEEE Int. Conf. Acoustics Speech and Sig. Proc. Proceedings}, 5.4, 346, 2006. 

 
\bibitem{NoMarkov2}
R.T.~Tranquillo and D.A.~Lauffenburger,
\newblock{Stochastic model of leukocyte chemosensory movement}
\newblock{\em J. Math. Biol.}, 25.5, 229, 1987. 


\bibitem{NoMarkov3}
E.~Gavagnin and C.A.~Yates,
\newblock{Stochastic and deterministic modelling of cell migration}
\newblock{\em Handbook of Statistics}, 39, 37, 2018. 

 

\bibitem{Abbas}
A.K.~Abbas and A.H.~Lichtman and S.~Pillai
\newblock{Basic immunology: functions and disorders of the immune system}
\newblock{\em Elsevier}, 2019. 


\bibitem{Agliari-SciRep2014}
E.~Agliari and E.~Biselli and A.~De Ninno  and G.~Schiavoni  and L.~Gabriele  and A.~Gerardino and F.~Mattei and A.~Barra and L.Businaro  
\newblock{Cancer-driven dynamics of immune cells in a microfluidic environment}
\newblock{\em Scientific Reports},  4.1, 6639, 2014. 

\bibitem{Biselli-SciRep2017}
E.~Biselli and et al.  
\newblock{Organs on chip approach: a tool to evaluate cancer -immune cells interactions}
\newblock{\em Scientific Reports},  7.1, 12737, 2017. 


\end{thebibliography}

\end{document}